\documentstyle[aps,epsf]{revtex}

\title{Basics of Quantum Computation}
\author{Vlatko Vedral and Martin B. Plenio \\
{\protect\small\em Blackett Laboratory, Imperial College, 
Prince Consort Road, London SW7 2BZ, U.K.}}

\date{\today}

\begin{document}
\maketitle

\begin{abstract}
Quantum computers require quantum logic, something fundamentally
different to classical Boolean logic. This difference leads
to a greater efficiency of quantum computation over its classical
counter--part. In this review we explain the basic principles 
of quantum computation, including the construction of
basic gates, and networks. We illustrate the power of
quantum algorithms using the simple problem of Deutsch, 
and explain, again in very simple terms, the well known algorithm 
of Shor for factorisation of large numbers into primes. We then 
describe physical implementations of quantum computers, 
focusing on one in particular, the linear ion--trap realization. 
We explain that the main obstacle to building an actual  
quantum computer is the problem of decoherence, which
we show may be circumvented using the methods of quantum 
error correction.      
\end{abstract}
\newpage
\tableofcontents
\newpage

\section{Introduction}

This review is not intended to cover all developments 
in the quantum information theory and quantum computation.
Our aim is rather to provide the necessary insights for 
an understanding of the field 
so that various non-experts can judge its fundamental
and practical importance.

Quantum computation is an extremely exciting and rapidly 
growing field of investigation \cite{AB96,EJ96,Steane98,PVK96,TMR98}. 
An increasing number of
researchers with a whole spectrum of different backgrounds, ranging from 
physics, via computing sciences and information theory to 
mathematics and philosophy, are involved in
researching properties of quantum--based computation. 
Interplay between mathematics and physics of course has always   
been beneficial to both types of human activities. 
The calculus was developed by Newton and Leibniz in
order to understand and describe dynamical laws of motion of material 
bodies. In general, geometry and physics have had a long and successful 
symbiotic relationship: classical mechanics and Newtonian
gravity are based on Euclidean Geometry, whereas in 
Einstein's Theory of General Relativity the basis is provided 
by non-Euclidean, Riemannian geometry, an important insight taken
from mathematics into physics. Although this link between Physics
and Geometry is still 
extremely strong, one of the most striking connections today is between
Information Theory and Quantum Physics and this will be investigated 
in the present review. 

Speaking somewhat loosely, we observe a trend 
to make mathematics ``more physical". What lies behind 
this phrase is the realization that the regularities and structures 
we observe in mathematics are actually deeply rooted in, and derive
from, the experiences of the physical world we happen to inhabit. 
According to this view, Geometry, for instance, does not have an independent, 
as it were Platonic, existence, but has to be inferred 
from making actual
measurements and observations in Nature. This thesis, that  
mathematics cannot be ``correct" {\em a priori}, but needs 
to be tested experimentally, was probably first fully 
realized by Einstein through General Relativity; the most recent 
example, however, is in the theory of computation. Computation, based on
the laws of classical physics,  leads  to completely different
constraints on information processing than computation based on quantum
mechanics (first realized by Feynman \cite{RPF82,RPF86} and Deutsch \cite{DD85}).  
This is an extraordinary fact:  we will show that
quantum information processing is faster, and, in some 
sense, more efficient than its classical counterpart (for a detailed
discussion of the physical basis of computation see \cite{DDBook}).  

Today's computers are classical, a fact which is actually 
not entirely obvious and is worth elaborating further. A basis of modern 
computers rests on semiconductor technology. 
Transistors, which are the ``neurons" of all computers,   
work by exploiting properties of semiconductors. However,
the explanation of how semiconductors function is
entirely quantum mechanical in nature: it simply 
cannot be understood classically. 
Are we thus to conclude
that classical physics cannot explain how classical computers
work?! Or are we to say that classical computers are, in fact, 
quantum computers! The answer to both these questions is yes and no. 
Yes, classical  
computers are in a certain, restricted, sense quantum mechanical, because,
as far as we understand today, everything is quantum mechanical.
No, classical computers, although based on quantum physics,
are {\em not} fully quantum, because they do not use
``quantumness" of matter at the information-theoretical level, where it
really matters.  Namely, in a classical computer information is
recorded in macroscopic, two level systems. Wires conducting
electrical current in computers can be in two basic states: 
when there is no current  
flowing through, representing a logical ``0", or else when there is 
some current flowing through, representing a logical ``1". These two states
form a {\em bit} of information. All computation is based on
logical manipulation of bits through logical gates acting on
wires representing these bits. 
Imagine, however, that instead of wires and currents 
we use two electronic states of an atom to record information. 
Let us call these states the ground state, $|0\rangle$, and the
excited state, $|1\rangle$ (Dirac notation is the most natural for quantum 
computing). But, since an atom obeys 
laws of quantum mechanics, the most general electronic state is
a superposition of the two basic states
\begin{equation}
|\Psi_1\rangle = a |0\rangle + b |1\rangle \; , 
\end{equation}
called the quantum bit or qubit, for short (this term was coined by
Schumacher \cite{BS95}). We see that in
addition to $0$ and $1$ states, a qubit has, so to speak, all the states 
``in between". When we have two bits, than there are four 
possibilities: $00,01,10,11$. However, this should be contrasted with
two qubits which are
in general in a state of the form
\begin{equation}  
|\Psi_{2}\rangle = a|00\rangle + b |01\rangle + c |10\rangle + d |11\rangle \; .
\end{equation}
If for example $a=d=0$ and $b=c=1/\sqrt{2}$, then we have
the famous Einstein-Podolski-Rosen (EPR) state \cite{Bell87}
\begin{equation}
|\Psi_{\mbox{EPR}}\rangle = \frac{(|01\rangle + |10\rangle)}{\sqrt{2}} \; .
\end{equation}
Two qubits in this state display a degree of correlation 
impossible in classical physics and hence violate the 
Bell inequality which is satisfied by all local  (i.e. classical) states. 
This phenomenon is 
called entanglement and is at the root of the success
of quantum computing. We will see how exploitation of 
a number of entangled qubits can lead to a considerable
computational speed-up in a quantum computer over its
classical counterpart.  Thus what distinguishes classical
and quantum computing is how the information is encoded
and manipulated, i.e. what plays a crucial role is whether the
{\em logical} basis is the classical, Boolean logic, or 
the quantum logic.

It is important to stress that apart from this 
theoretically--driven curiosity to investigate
quantum computation, there is a practical need to do so, too.
This stems from the observation of the rate of technological
progress which is known as Moore's Law. This law states that 
the number of transistors per
chip (i.e. the complexity of computers) grows  
exponentially with time; more precisely, it doubles 
every year, as depicted in Fig. 1 (the 1997 special issue of
Scientific American is devoted entirely to the technological side of
computation \cite{SA}, including an article about Moore's Law). 
We can see that this law has been obeyed almost precisely in the last
$30$ years. If this exponential growth is 
extrapolated into the near future we see that, at Moore's rate, a bit of 
information will be encoded into a single atom by the year 2017. 
In fact, even before that, by the year 2012, quantum effects will 
become very important, so much so that they will have a considerable
effect upon computation and should not be neglected any longer. 
Thus, not only our theoretical
curiosity, but also technological progress requires 
that we study and understand quantum computation.

\section{Computation in Classical and Quantum Physics}

We have noted that classical computation is based on Boolean
logic. This logic can be represented in terms of gates
acting on particular bits of information. In fact, a classical 
computer can be 
viewed as a collection of bits and gates which act on
a certain input, i.e. the initial state of bits, to produce
via the gate action a certain output, i.e. the final state
of bits. We will review some properties of classical
gates in the next subsection, and then see how this picture differs
in the case of a quantum computer.

\subsection{Classical Gates}

Let us start with the simplest of logical gates, i.e. one bit gates.
What can happen to a single bit? The first possibility is ``nothing", 
but this is a 
trivial gate which of course does not alter the state of the bit.
Secondly, the value of the bit $0$ can turn into $1$ and $1$ into $0$, 
which is called a NOT gate. This exhausts all the possibilities
for a single bit. 

Two bits are much more exciting. A usual two bit gate has two
input bits and one output bit. Take, for example, an OR gate:
the output of this gate is $0$ only if both the input bits are 
$0$, and otherwise it is $1$ (see Fig. 2.). This gate illustrates an important
property of classical computers, all of which contain some OR gates.
Namely, the OR gate is irreversible, meaning that given the
value of the output bits we cannot reconstruct the values of the input bits.  
So, if the output to the OR gate was $1$, the input could 
have been $11$, $01$ or $10$, i.e. it is simply undetermined.   
Thus this gate cannot really be run backwards, and hence 
is called irreversible. 

Now, how many two--input--one--output--bit gates are there? 
Well, there are $4$ possible different inputs and each one can have 
a different output leading to $4^2=16$ possible different gates. 
However, not all the gates
are necessary: there are sets of gates, called fundamental 
gates, out of which any other gate can be constructed. For
example, using NOT, OR and AND gates we can construct any
other gate. A more striking example is a SHEFFER gate out 
of which any other gate can be constructed as proven by 
Sheffer \cite{Sheffer}. 
If $p$ and $q$ are two bits assuming either $0$ or $1$ values, i.e.
$p$ and $q$ are two {\em binary variables}, then 
this gate can be written as NOT ($p$ OR $q$), as in Fig. 2. 
Since any computation,
i.e. any combination of any gates on any number of bits, can be
written in terms of Sheffer gates only, this gate is called {\em universal}. 

Another interesting feature of computers based on irreversible 
gates (i.e. all existing real world computers) is that they 
dissipate energy as they run \cite{RL61}. 
This energy is actually dissipated in the form of
heat when the information 
is {\em deleted} in the irreversible gates we described 
above, as discovered by Landauer \cite{RL61}. In fact,    
deleting information always involves investing work and
wasting energy \cite{CHB73}, as can be illustrated in the following 
simple example (for a detailed discussion at a simple level see 
\cite{RPF97}). Consider a container enclosing a single
gas atom, as in Fig. 3 a). If the atom is in the left hand half of the  
container, we take this to represent a logical $0$,
and if the atom is in right hand half of the container
we take this to represent a logical $1$. Say that initially
we do not know where the atom is. Since there are two
possibilities for the state of the atom, the entropy of
the atom is equal to $k \ln 2$, where $k$ is the Boltzmann
constant (this relationship between thermodynamical entropy and
information, or uncertainty, was first explicitly emphasised by
Shannon in his celebrated Information Theory \cite{SW48}, although
it was Szilard who hinted at this relationship much earlier \cite{Szilard}). 
Deleting this information means finding out what 
the state of the atom is, i.e. 
confining the atom to the particular (known) 
half of the container. To do this, we can, for example, push 
the right hand wall to the left until the atom is
confined entirely to the left hand half, as in Fig. 3 b). If we
do this at a constant temperature $T$ (most 
computers work at room temperature anyway!) 
we get a heat loss of $kT \ln 2$ to the environment. Now the same
effect is present in any computer and that is why they
heat up as they work (in fact, they use much more
than $kT\ln 2$ per gate: ordinary PCs use about $10^8 kT$ per gate,
but $kT\ln 2$ is the fundamental thermodynamical limit, 
as first shown by Landauer \cite{RL61}). 

The reader should be 
warned, however, that the above example is deceptively simple.
The above process can be thermodynamically reversible,
in which case the increase in the entropy of environment is
compensated by the decrease in the entropy of the atom in
the box. In computation, on the other hand, logically
irreversible operations are performed on deterministic data, 
so that the increase in the entropy of the environment is not
compensated by a decrease in the entropy of the system, and
hence the operation is also thermodynamically irreversible.
The question that is then important from the fundamental 
point of view is whether we can manage without this: can 
we run computers without any energy dissipation? 
Surprisingly, the answer is yes! Instead of irreversible 
gates we have to use
reversible gates; it can be shown that reversible gates 
can perform without any heat loss \cite{CHB73}. In addition, Bennett
proved that any irreversible computation can be performed
using reversible gates only \cite{CHB73}. So, the crucial question is how to
make reversible gates out of irreversible gates.

Let us illustrate the general principle using the
OR gate. We emphasised that this gate is irreversible
because there is only one output and two inputs, so that
information gets lost in the gate. So, we can decide
to add another output bit which, for example, 
``saves" the value of the first input bit as in Fig. 4. 
Thus, with two inputs and two outputs the resulting
gate is reversible. This is, in fact, the general method
of constructing reversible out of irreversible gates:
we have to save as much of the input as necessary at the 
output, so that, given that output, we can unambiguously
determine the value of the input. In this way all the
irreversible computation can be made reversible 
and, at least in principle, dissipation--free. We stress again that
logical reversibility is just a necessary condition for
no heat loss, and we have to make the computation thermodynamically
reversible as well. Bennett constructed
thermodynamical models of computers which dissipated arbitrarily
little energy if run sufficiently slow \cite{CHB73}, and
his review of the field of reversible computation is 
given in \cite{CHB88}.  An interesting model of a 
billiard ball reversible computer was also developed by
Fredkin and Toffoli \cite{FT82}. We
say that these models are reversible ``in principle", 
because we know that thermodynamical
reversible processes are quasi-static idealizations 
of real processes, which are never exactly true under
realistic circumstances.

We will see that this question of reversible computation 
is not only relevant to heat--free
classical computation, but also is of central 
importance for quantum computing. Next we 
explain how.

\subsection{Quantum Gates}

For completeness let us start with some basic definitions. A quantum
network is a quantum computing device consisting of quantum logic
gates whose computational steps are synchronised in time 
\cite{DD85,RPF82,RPF86}. The outputs
of some of the gates are connected by wires to the inputs of others.
The size of the network is governed by its number of gates. 
The size of the input
of the network is governed by its number of input qubits {\em i.e.\/} the qubits
that are prepared appropriately at the beginning of each computation
performed by the network. Inputs are encoded in binary form in the
computational basis of selected qubits often called a {\em quantum
  register\/}, or simply a {\em register\/}. For instance, the binary
form of the number $6$ is $110$ and loading a quantum register with this
value is done by preparing three qubits in the state $|1\rangle \otimes
|1\rangle \otimes |0\rangle$.  In the following we use a more compact
notation: $|a\rangle$ stands for the direct product $|a_n\rangle
\otimes |a_{n-1}\rangle \ldots|a_1\rangle \otimes|a_0\rangle$ which
denotes a quantum register prepared with the value $a=2^0 a_0+ 2^1 a_1
+\ldots 2^n a_n$. Computation is defined as a unitary evolution of the
network which takes its initial state ``input'' into some final state
``output'' (analogous to classical computation).

The entire quantum computation is thus a unitary transformation,
where a measurement is performed at the end to extract the result.
This will be explained in more detail in the next subsections
through a couple of examples. However, a unitary transformation
is itself reversible; therefore, we have to use reversible
gates introduced previously in order to be able to implement
quantum gates. The difference between reversible and quantum
computation is that a quantum gate acts on superpositions of
different basis states of qubits, whereas classically 
this option is non--existent. In Fig. 5 we present three 
basic gates used in quantum computation, the NOT gate,
the Controlled NOT gate and the TOFFOLI gate \cite{ABC95}. Controlled
NOT gate (CNOT, for short) is a two qubit gate, where the value of
the first qubit (called control) determines what will happen to the 
second qubit (called target) qubit. Namely if the control qubit is $1$, 
we apply the NOT gate to the target qubit and otherwise nothing happens 
to it (hence the name Controlled NOT). TOFFOLI gate can be understood 
as Controlled--Controlled NOT. As in
classical computation, there are universal gates in
quantum computation. There is, for example, a three qubit 
gate which is universal, discovered by Deutsch \cite{DD89}, 
and also a two qubit gate
which is universal (e.g. \cite{AB95}). An extremely useful result 
of this universality is that
any quantum computation can be done in terms of 
a Controlled NOT gate and a single qubit gate \cite{DBE95,SL95} (which varies), 
although,
of course, it might sometimes be more convenient to use
other gates as well \cite{AB95}. An important one qubit gate is 
the so called Hadamard transformation whose action is 
the following (the normalization is omitted)
\begin{eqnarray}
|0\rangle & \longrightarrow &  |0\rangle + |1\rangle \\
|1\rangle & \longrightarrow &  |0\rangle - |1\rangle
\end{eqnarray}      
This transformation will be used frequently throughout this review.

Both the input and the output of a quantum computer 
can be encoded in several registers.
Even when $f$ is a one--to--one map between the input $x$ and the
output $f(x)$ and the operation can be formally written as a unitary
operator $U_f$
\begin{equation}
  U_f|x\rangle\rightarrow |f(x)\rangle,
\label{bij}
\end{equation}
we may still need an auxiliary register to store the intermediate
data. When $f$ is not a bijection we definitely {\em have} to use an additional
register in order to guarantee the unitarity (i.e reversibility)
of computation. In this
case the computation must be viewed as a unitary transformation $U_f$
of (at least) two registers
\begin{equation}
  U_f |x,0\rangle \rightarrow |x,f(x)\rangle,
\end{equation}
where the second register is of appropriate size to accommodate
$f(x)$. This uses the same principle demonstrated in making the OR
gate reversible. We now show a simple, but extremely important use
of the Controlled NOT gate.

\subsection{Quantum Entanglement}

We stressed in the introduction that quantum entanglement
is the phenomenon responsible for all the advantages of 
quantum computation. Here we show how to create entangled 
quantum states using the simple quantum gates introduced
previously. This operation is used time and again in various
quantum computations as we will see when reviewing Shor's
quantum algorithm for factorization of natural numbers into 
primes \cite{PWS94}. 

To illustrate entanglement we look at the EPR-state of two
qubits 
\begin{equation}
|\Psi_{\mbox{EPR}}\rangle = \frac{(|01\rangle + |10\rangle)}{\sqrt{2}} \; .
\end{equation}
We say that a pure state of two qubits is entangled if
it cannot be written as a product of the individual states
of the two qubits, such as $|\psi_1 \rangle \otimes |\psi_2\rangle$. 
The EPR state is obviously not decomposable into a direct 
product of any form, and is therefore entangled. 
Of course, both of the states $|01\rangle$ and $|10\rangle$ are
of the direct product form, but their superposition is not. 
The interesting question is therefore how to create an 
EPR state starting from just a disentangled, say, the $|01\rangle$ 
state. The required quantum computation is very simple: 
first we apply a Hadamard transformation to
the first qubit, and then a Controlled NOT between the first 
qubit and the second qubit, where the second qubit is the target.  
These two steps can be written as (the normalization is omitted)
\begin{eqnarray}
|01\rangle \longrightarrow (|0\rangle + |1\rangle)|1\rangle \longrightarrow 
|01\rangle + |10\rangle \; .
\end{eqnarray}    
We see that after the action of the Hadamard transformation the qubits are
still disentangled. This is because this transformation acts on only
one of the qubits, i.e. is applied {\em locally} and not {\em globally}, 
and therefore
cannot create global features such as entanglement. 
This is true, in general, implying that no local operation
whatsoever can create an entangled state out of a disentangled
one, a principle which has a fundamental place in quantum information
processing (see e.g. \cite{Bennett96,VPRK97,VPJK97,VRP97,VP98}). 
Only global transformations such as a Controlled NOT can 
create entanglement. 

The above was the first and simplest form of quantum computation
involving only two qubits and a few gates. Let us now 
look at some slightly more complicated examples.

\section{Simple Quantum Networks}

In this section we present some very simple quantum networks.
These networks will provide a basis for the more complicated
Shor's algorithm \cite{PWS94} reviewed in the next section.  

\subsection{Simple Arithmetic}

Quantum networks for basic arithmetic operations can be constructed in
a number of different ways. Although almost any non-trivial quantum
gate operating on two or more qubits can be used as an elementary
building block of the networks~\cite{AB95} we have decided to use the
three gates described in Fig.~\ref{basicgates}, hereafter referred to
as {\em elementary gates}. None of these gates is universal for
quantum computation; however, they suffice to build any Boolean
functions as the Toffoli gate alone suffices to support any {\em
  classical} reversible computation \cite{TT81}. The {\small\sf NOT} and the
Control--{\small \sf NOT} gates are added for convenience (they can be
easily obtained from the TOFFOLI gates, an exercise we leave to the reader).

\subsection{Plain adder}

The addition of two registers $|a\rangle$ and $|b\rangle$ is probably
the most basic arithmetic operation \cite{VBE96}. In the simplest form 
it can be written as
\begin{equation}
  |a,b,0\rangle \rightarrow |a,b,a+b\rangle.
\end{equation}
Here we will focus on a slightly more complicated (but more useful)
operation that rewrites the result of the computation into the one of
the input registers,  which is the usual way additions are 
performed in conventional irreversible hardware; {\em i.e.\/}
\begin{equation}
  |a,b\rangle \rightarrow |a,a+b\rangle,
\end{equation}
As one can reconstruct the input $(a,b)$ out of the output $(a,a+b)$,
there is no loss of information, and the calculation can be
implemented reversibly. To prevent overflows, the second register
(initially loaded in state $|b\rangle$) should be sufficiently large,
{\em i.e.\/} if both $a$ and $b$ are encoded on $n$ qubits, the second
register should be of size $n+1$. In addition, the network described
here also requires a temporary register of size $n-1$, initially in
state $|0\rangle$, to which the carries of the addition are
provisionally written (the last carry is the most significant bit of
the result and is written in the last qubit of the second register).

The operation of the full addition network is illustrated in
Fig.~\ref{plainadder} and can be understood as follows:
\begin{itemize}
\item
We compute the most significant bit of the result $a+b$. This step
requires computing all the carries $c_i$ through the relation $c_i
\leftarrow a_i$ {\small \sf AND} $b_i$ {\small \sf AND} $c_{i-1}$,
where $a_i$, $b_i$ and $c_i$ represent the $i$th qubit of the first,
second and temporary (carry) register respectively. Fig.~\ref{carrysum}i)
illustrates the sub--network that effects the carry calculation. 
\item Subsequently we reverse all these operations (except for the
  last one which computed the leading bit of the result) in order to
  restore every qubit of the temporary register to its initial state
  $|0\rangle$. This enables us to reuse the same temporary register,
  should the problem, for example, require repeated additions. During
  the resetting process the other $n$ qubits of the result are
  computed through the relation $b_i \leftarrow a_i$ {\small \sf XOR}
  $b_i$ {\small \sf XOR} $c_{i-1}$ and stored in the second register, where
  XOR is the Exclusive OR gate: the output is $0$ if the inputs are $0,0$ or
  $1,1$, and otherwise is $1$.
  This operation effectively computes the $n$ first digits of the sum
  (the basic network that performs the summation of three qubits
  modulo $2$ is depicted in Fig.~\ref{carrysum}ii).)
\end{itemize}

The addition network has a typical "V" shape present in any reversible
computation which has to dispose of unneccessary information (i.e. garbage).
We show in the next section why this is the case.
Another interesting feature of the adder is that if we reverse the action
 of the above network ({\em i.e.\/} if we
apply each gate of the network in the reversed order) with the input
$(a,b)$, the output will produce $(a,a-b)$ when $a\geq b$. So, with
the same network we can also accomplish subtraction! When $a<b$,
the output is $(a,2^{n+1}-(b-a))$, where $n+1$ is the size of the
second register. In this case the most significant qubit of the second
register will always contain $1$ . By checking this ``overflow bit''
it is therefore possible to compare the two numbers $a$ and $b$; we
can use this operation to construct the network for modular addition.
The crucial fact is that once we know how to perform a modular addition, 
we immediately 
know how to execute modular multiplication \cite{VBE96}, since
\begin{equation}
a\times b \,\mbox{mod} N = (\underbrace{a + a \ldots + a}_{b\, \mbox{times}}) \,\mbox{mod} N \; .
\end{equation}
Likewise we can perform modular exponentiation \cite{VBE96}, because
\begin{equation}
a^x \mbox{mod} N = (\underbrace{a \times a \ldots \times a}_{x\, \mbox{times}}) \,\mbox{mod} N \; .
\end{equation}
Thus addition is at the root of all the other simple arithmetic 
operations: subtraction, (modular) multiplication and (modular) exponentiation.
It should be noted that multiplication (and therefore exponentiation) can
be performed more efficiently than this (see the fastest multiplication 
algorithm in \cite{SS71}), however the above will be sufficient for our 
purposes.  Modular operations will become particularly important
when we discuss Shor's algorithm shortly \cite{VBE96}.  Next, however,
we consider another important problem, i.e. that of {\em garbage reduction}.

\subsection{Garbage Disposal}

We have seen that in order to perform more involved 
arithmetic operations we need to repeat the simpler
ones a number of times. However, each of these 
simple operations contains a number of additional,
auxiliary qubits, which serve to store the intermediate
results, but are not relevant at the end. In order
not to waste any unneccesary space, it is therefore
important to reset these qubits to $0$ so that we are
able to re--use them. A good example of a procedure
which does not generate any garbage is the 
reversible addition we introduced earlier; all the
carry bits were reset to $0$ at the end of addition.
Quantum mechanically, this reset is even more
important. This is because the result of
any quantum computation is always entangled with
the auxiliary, garbage qubits. Consider the output of the 
adder network, where the carry qubits are auxiliary. 
It will be in a superposition
of different results $|r_i\rangle$ each one 
entangled to a different input $|a_i\rangle$
and carry $|c_i\rangle$, 
\begin{equation}
\sum_i |r_i\rangle\otimes |a_i\rangle \otimes |c_i\rangle \; .
\end{equation}
For clarity let us consider two terms only,
\begin{equation}
|r_1\rangle\otimes |a_1\rangle \otimes |c_1\rangle + |r_2\rangle\otimes |a_2\rangle \otimes |c_2\rangle\; .
\end{equation}
If the garbage is now reduced to $|0\rangle$ then the total 
state becomes
\begin{equation}
(|r_1\rangle\otimes |a_1\rangle+ |r_2\rangle\otimes |a_2\rangle )\otimes |0\rangle \; ,
\end{equation}
i.e. the garbage is disentangled from the rest. If, however, this is
not done, and we completely disregard the state of the carries in the
further computation then the effective state of the rest of 
result is obtained by tracing over the states of $|c\rangle$
to obtain,
\begin{equation}
|r_1\rangle\langle r_1|\otimes |a_1\rangle\langle a_1| + |r_2\rangle \langle r_2| \otimes |a_2\rangle\langle a_2| \; .
\end{equation} 
This state is now a mixed state and is completely disentangled. We
have already said that as soon as entanglement disappears,
then a quantum computer is no more powerful than a classical computer.
Thus resetting the garbage plays a central importance in
quantum computation (classical reversible computation does
not suffer from this since there is no entanglement in the first place!). 
More generally, a quantum computer will interact 
with its environment and become entangled with it in 
exactly the same way that the carries and the other two registers become
entangled.
However, the environment is in general impossible to control
and this means that the pure superposition states of a quantum 
computer eventually become mixed (and therefore useless from the
quantum computational point of view). This is the most general model
of how errors arise in quantum computation, and will be described in
more detail later. Here we first present a universal way 
of dealing with ``garbage bits" that arise during a computational 
process.  

Suppose that we have to compute $f^4(x)=f(f(f(f(x))))$ given $x$ and
given that we know how to reversibly perform
\begin{equation}
|x\rangle |0\rangle \longrightarrow |x\rangle |f(x)\rangle \;\; .
\end{equation}   
A naive way would be to prepare five registers in the state
$|x\rangle |0\rangle |0\rangle |0\rangle |0\rangle$ and then to execute 
computation of $f$ four times
\begin{eqnarray}
|x\rangle |0\rangle |0\rangle |0\rangle |0\rangle & \longrightarrow &
|x\rangle |f(x)\rangle |0\rangle |0\rangle |0\rangle \\
&\longrightarrow  & |x\rangle |f(x)\rangle |f(f(x))\rangle |0\rangle |0\rangle\\
&\longrightarrow  & |x\rangle |f(x)\rangle |f(f(x))\rangle |f(f(f(x))) \rangle |0\rangle\\ 
&\longrightarrow  & |x\rangle \underbrace{|f(x)\rangle |f(f(x))\rangle |f(f(f(x))) \rangle}_{\mbox{``garbage"}} |f(f(f(f(x))))\rangle \;\; .
\end{eqnarray}
However, we have now generated three (middle) unwanted, garbage registers (see Fig. 8).
We were only interested in computing $f^4$ and we do not need all the
intermediate results. So, how do we re--set them to zero?
We simply add another register and bitwise copy our result into it (this is
accomplished by a CNOT);
then we reverse the whole computation ending up with
\begin{eqnarray}
|x\rangle |0\rangle |0\rangle |0 \rangle |0\rangle |f(f(f(f(x))))\rangle \; .
\end{eqnarray}
This trick can be improved even further \cite{CHB89,LS90}, but these improvements
are too detailed to be relevant here. The whole issue 
of reversibility becomes extremely important in constructing a network
for Shor's algorithm. However, before we
analyse this, the biggest breakthrough in quantum computing we turn to
a simpler example, Deutsch's problem \cite{AE97}, which illustrates all the
basic properties of quantum algorithms.

\subsection{Deutsch's Problem}

Deutsch's problem is the simplest possible example which illustrates 
the advantages of quantum computation through exploiting
entangled states.  The problem is the following. 
Suppose that we are given a binary function of a
binary variable $f: \{ 0,1\} \longrightarrow \{ 0,1\}$.  
Thus, $f(0)$ can either be $0$ or $1$, and $f(1)$ likewise
can either be $0$ or $1$, giving altogether four possibilities. 
However, suppose that we are not interested in the particular
values of the function at $0$ and $1$, but we need to 
know whether the function is: 1) constant, i.e. $f(0)=f(1)$, or
2) varying, i.e. $f(0)\ne f(1)$. Now Deutsch poses the following
task: by computing $f$ only {\em once} determine whether it is   
constant or varying. This kind of problem is generally referred to as
a {\em promise algorithm}, because one property 
out of a certain number of properties 
is initially promised to hold, and our task is to determine 
computationally which one holds (see also \cite{DJ92,DS94} for other
similar types of promise algorithms).   

First of all, classically this is clearly impossible. We
need to compute $f(0)$ and then compute $f(1)$
in order to compare them. There is no way out of this
double evaluation.  
Quantum mechanically, however, there is a simple  
method to achieve this task by computing $f$ only once! 
Two qubits are needed for the computation. 
We can imagine that the first qubit is the input
to the quantum computer whose internal (hardware)
part is the second qubit. The computer itself will 
implement the following transformation on the 
two qubits 
\begin{equation}
|x\rangle |y\rangle \longrightarrow |x\rangle |y \oplus f(x)\rangle \; ,  
\label{dd}
\end{equation}
where $x$ is the input qubit and $y$ the hardware, as depicted in Fig. 9.
Note that this transformation is reversible and thus there is a
unitary transformation to implement it (but we will not pay any attention 
to that
at the moment, as we are only interested here 
in the basic principle). Note also   
that $f$ has been used only once. The trick is to prepare the 
input in such a state that we make use of quantum entanglement.
Let us have at the input
\begin{equation}
|x\rangle |y\rangle = (|0\rangle + |1\rangle)(|0\rangle - |1\rangle)\; ,
\label{in}
\end{equation}
where $|x\rangle$ is the actual input and $|y\rangle$ is part
of the computer hardware.
Thus before the transformation is implemented, the state
of the computer is in an equal superposition of all four
basis states, which we obtain by simply expanding the state in eq. (\ref{in}),
\begin{equation}
|\Psi_{\mbox{in}}\rangle = |00\rangle - |01\rangle + |10\rangle - |11\rangle\; .
\end{equation}
Note that there are negative phase factors before the second and fourth term.
When this state now undergoes the transformation in eq. (\ref{dd}),
we have the following output state
\begin{eqnarray}
|\Psi_{\mbox{out}}\rangle & = &  |0f(0)\rangle - |0\overline{f(0)}\rangle + |1f(1)\rangle - 
|1\overline{f(1)}\rangle \\
& = &  |0\rangle (|f(0)\rangle - |\overline{f(0)}\rangle) + |1\rangle 
(|f(1)\rangle -  |\overline{f(1)}\rangle ) \; ,
\end{eqnarray}
where the bar indicates the {\em opposite} of that value, 
so that, for example, $\overline{0}=1$.
This is an entangled state and now we see where the power of quantum 
computers is fully realised: each of the components in the
superposition of $|\Psi_{\mbox{in}}\rangle$ underwent the  
same evolution of eq. (\ref{dd}) ``simultaneously", leading to
the powerful ``quantum parallelism" \cite{DD85}. Let us look at the two 
possibilities now.
\begin{enumerate}
\item if $f$ is constant then  
\begin{equation}
|\Psi_{\mbox{out}}\rangle = (|0\rangle + |1\rangle) (|f(0)\rangle - 
|\overline{f(0)}\rangle ) \; .
\end{equation}
\item if $f$ is varying then 
\begin{equation}
|\Psi_{\mbox{out}}\rangle = (|0\rangle - |1\rangle) (|f(0)\rangle - 
|\overline{f(0)}\rangle ) \; .
\end{equation}
\end{enumerate}
Note that the output qubit (the first qubit) emerges in two 
different orthogonal states, depending on the type of $f$.
These two states can be distinguished with $100$ percent
efficiency. This is easy to see if we first perform a
Hadamard transformation on this qubit, leading to
the state $|0\rangle$ if the function is constant,
and to the state $|1\rangle$ if the function is 
varying. Now a single projective measurement in
$0,1$ basis determines the type of the function.
Therefore unlike their classical equivalents
quantum computers can solve Deutsch's
problem.

It should be emphasised that this quantum computation, 
although extremely simple, contains all the main features 
of successful quantum algorithms: it can  
be shown that all quantum computations are just
more complicated variations of Deutsch's problem \cite{CEMM97}. 
Deutsch's algorithm has now been implemented 
successfully using nuclear magnetic resonance methods \cite{Jones97,Gershenfeld97}.
Of course, the more complicated the computation, the
more qubits involved, and the easier it is to see the
difference between quantum and classical computations, especially
the greater difference in their efficiency. 
The most striking example to date is Shor's
algorithm for the factorization of large numbers into 
primes, which we review next (there are other examples 
of quantum computation being faster than their classical counterpart, 
see e.g. \cite{LKG97}, however none of them is as decisive as
Shor's algorithm).

\section{Outline of Quantum Factorization}

The algorithm for factorization dates back to the Ancient 
Greeks (the book by Knuth in \cite{DEK81} is a bible 
for algorithms, containing a number of important 
classical computational problems). It was probably 
known to Euclid, and it can be
described simply as follows. We wish to find the prime 
factors of $N$. This amounts to finding the smallest 
$r$ such that $a^r \equiv 1 (\mbox{mod} N)$, where $a$ is chosen
to be coprime to $N$, i.e. so that $a$ and $N$ have no common
divisors apart from $1$. In other words, we want to determine the
period of the function $a^r (\mbox{mod} N)$. Let us see how this works for, 
say, $N=15$.
\begin{itemize}
\item We choose a=2. Then obviously $\gcd (2,15) =1$. 
\item Next we compute $2^0,2^1,\ldots 2^i$ modulo $15$, and this  
gives $1,2,4,8,1,2,4,8,\ldots$.
\item This sequence is periodic with the period $r=4$,
which also satisfies $2^4 \equiv 1 (\mbox{mod} 15)$.
\item Once $r$ is obtained we find the factors of N by
computing $\gcd (a^{r/2} \pm 1,15)$, which in our case is
$\gcd (4\pm 1,15) =3,5$.
\end{itemize}
Hence we have factorised $15$ into $3\times 5$. Now this algorithm
(or some of its close variants) can be implemented on a 
classical or on a quantum computer. To be able to compare 
their efficiency we need to know that there are two basic
classes of problems:
\begin{enumerate}
\item {\em easy problems}: the time of computation $T$ is a polynomial
function of the size of the input $l$, i.e. $T=c_n l^n + ...+ c_1 l+c_0$, 
where the coefficients $c$ are determined by the problem.
\item {\em hard problems}: the time of computation is an exponential 
function of the size of the input (e.g. $T = 2^{cl}$, where c is problem
dependent).
\end{enumerate}
The size of the input is always measured in bits (qubits). For
example, if we are to factorize $15$ then we need $4$ bits 
to store this number. In general, to store a number N we
need about $l=\log N$, where the base of the logarithm is $2$.  
(this is easy to see: just ask yourself how many different numbers 
can be written with $l$ bits).  The easy problems are considered
as computationally efficient, or tractable, whereas the
hard problems are computationally inefficient, or intractable.
Now the upshot 
of this discussion is that, for a given $N$, 
there is no {\em known} efficient classical algorithm to
factorise it. Let us illustrate how the simplest factorization
algorithm performs: suppose that we want to determine the
factors of $N$ by dividing it by $2$, then $3$ then $4$ and
so on up to $\sqrt{N}$. So the time of computation (which is
in fact the number of elementary steps) is proportional 
to the number of divisions we have to perform, and this is
$\sqrt{N} = 2^{l/2}$, i.e. it is exponential.  
However, using a quantum computer 
and the above--described Euclid's approach to factorization, we
can factor any $N$ efficiently in polynomial
time involving a linear number of qubits. 
This is essence of Shor's algorithm.

There are two distinct stages in this algorithm \cite{PWS94}
(for an extensive review
of this algorithm see \cite{EJ96}).
Initially, we have two registers (plus several other registers 
containing garbage, but these are
irrelevant for explaining the basic principle of quantum factorization)
at the input to the quantum
computer. First, we prepare the first register in a  
superposition of consecutive natural numbers, while leaving
the second register in $0$ state to obtain  (as usual
we omit the normalization)
\begin{equation}
|\Psi \rangle = \sum_{n=0}^{M-1} |n\rangle |0\rangle 
\end{equation}
where $M=2^m$ is some sufficiently large number.
Now in the second register we compute the function
$a^i \mbox{mod} N$. This can be achieved unitarily 
and the result is
\begin{equation}
|\Psi_1 \rangle = \sum_{n=1}^{M-1} |n\rangle |a^n \mbox{mod} N\rangle \; .
\label{exp}
\end{equation}
Here again we see famous quantum parallelism in action.  
This completes the first stage of the algorithm and
the trick now is to extract the period $r$ from 
the first register. To help us visualize this
let us think of our previous example when $N=15$ and
$a=2$. Then we would have
\begin{eqnarray}
|\Psi_1 \rangle & = & |0\rangle |2^0 \mbox{mod} 15\rangle + |1\rangle 
|2^1 \mbox{mod} 15\rangle + |2\rangle |2^2 \mbox{mod} 15\rangle + |3\rangle 
|2^3 \mbox{mod} 15\rangle + \nonumber \\
& + & |4\rangle |2^4 \mbox{mod} 15\rangle + |5\rangle |2^5 \mbox{mod} 15\rangle \ldots + |2^{M-1} \mbox{mod} 15\rangle\\  
& = & |0\rangle |1\rangle + |1\rangle 
|2\rangle + |2\rangle |4\rangle + |3\rangle 
|8\rangle + |4\rangle |1\rangle + |5\rangle |2\rangle \ldots  
+ |2^{M-1} \mbox{mod} 15\rangle\; .
\end{eqnarray}
Let us recall that we do not need to extract 
all the values of $2^i \mbox{mod} 15$, but
just the period of this function.  This now sounds very
much like Deutsch's problem, where only the knowledge of 
a property of $f$ was important and not both its values. The
solution is likewise similar, but is however much more
computationally involved. Suppose that we now perform a measurement 
on the second register to determine its state. Suppose 
further that we obtain 4 as the result. The remaining 
state will be 
\begin{equation}
|\Psi_2 \rangle =  (|2\rangle + |6\rangle + |10\rangle + \ldots)|4\rangle
\end{equation}
so that the first register contains numbers repeating periodically
with the period $4$. This is now what we have to extract by
manipulating the first register. To see how this works 
suppose for simplicity that $r$ divides $M$ exactly. 
For general $a$ and $N$ this state is 
\begin{equation}
|\Psi_2 \rangle = \sum_{j=0}^{A} |jr + l\rangle |l\rangle
\end{equation}
where $A=M/r-1$ and the second register is obviously irrelevant.
Extracting $r$ involves performing a Fast Fourier Transform  
on the first register, so that the final state becomes
\begin{equation}
|\Psi_3 \rangle = \sum_{j=0}^{r-1} \exp (2\pi ilj/r) |jM/r\rangle 
\end{equation}
We can now perform a measurement in the $y=jM/r$ basis where $j$ is an
integer. Therefore, once we obtain a particular $y$ we have 
to solve the following equation $y/M=j/r$ where $y$ and $M$ are 
known.  Assuming that $j$ and $r$ have no common factors (apart
from $1$) we can determine $r$ by cancelling $y/M$ down to 
an irreducible fraction. Once $r$ is obtained we can easily  
infer the factors of $N$.

In general, of course, Shor's algorithm is probabilistic.  
This means that $r$, and hence the factors of $N$ that we obtain
by running the above quantum computation, might sometimes
not be the right answer. 
However, whether the answer is right or wrong can be 
easily checked by multiplying the factors to get $N$. 
Since multiplication is an easy computation this can be  
performed efficiently on a classical computer. 
If the result is not $N$, we then
repeat the whole Shor's algorithm all over again, and we 
keep doing this until we get the right answer. Shor
showed that even with this random element his algorithm
is still efficient. In fact, the most time consuming 
part is the first one, where we have to obtain the state   
in eq. (\ref{exp}). Modular exponentiation takes of 
the order of $(\log N)^3$ elementary gates and this 
dominates the whole algorithm \cite{VBE96}. We should say that  
the memory space, i.e. the number of qubits needed
for the entire computation, is of the order of $\log N$.
For completeness we state that all the above networks 
for addition, multiplication and exponentiation can be 
improved using standard computational techniques (see e.g. \cite{divide}),
however, this improvement is not substantial and does not 
change the fundamental conclusion about the efficiency
of quantum factorization.

All the computations we have considered thus far are ``ideal"
in the sense that every gate operation was assumed to be performed 
without error, and in addition states of qubits were completely 
preserved. In reality this is far from being the case. Quantum 
information is usually very fragile and easily undergoes errors 
deriving from the so called phenomenon of decoherence. Manipulating 
quantum information also presents a formidable experimental task. 
This is the subject of the remainder of our review. 

\section{How to realize a Quantum Computer}

In the previous sections we have seen that quantum computation is a 
fundamentally new concept that promises the solution of
problems which are intractable on a classical computer. In this 
section we will now address the question of how to implement such
a quantum computer in practise. The interest in the practical realization 
of a quantum computer increased substantially 
after the discovery of Shor's factorization algorithm \cite{PWS94}, 
which exhibited
the great potential of quantum computation. An important question 
was whether a quantum computer required fundamentally new experimental
techniques for its realization, or whether already known techniques would
be sufficient. In fact, some of the early proposals for implementations 
of quantum computers had the disadvantage
of using somewhat 'futuristic' experimental techniques. Then, however, a very
beautiful proposal for an ion-trap quantum computer was made by Cirac and 
Zoller \cite{Cirac95} which employed only experimental methods which 
were already realized or which were expected to be realizable in the near 
future. Subsequently, other realistic suggestions such as quantum computation 
based on nuclear magnetic resonance methods have been made 
\cite{Cory97,Gershenfeld97,Knill97b,Laflamme97}. 
Although these new proposals are very interesting, we confine 
ourselves here to the description of the linear ion trap implementation
of Cirac and Zoller. The reason for this is that the ion trap quantum computer
exhibits the basic ideas of quantum computation in a particularly transparent
and beautiful way.

\subsection{The ion trap quantum computer}
The basic experimental setup for an ion trap quantum computer is given in 
Fig. \ref{lintrap}. In a linear ion trap ac currents in the electrodes 
generate a time dependent electric quadrupole field. Ions move in this 
potential and for suitable ac currents and ion masses, the ions are trapped.
This means that they see an effective force towards the center of the 
axis of the linear ion trap. The equilibrium between the trapping force 
generated by the electrodes and the mutual electrostatic 
repulsion of the ions is given when the ions form a string where adjacent 
ions are separated by a few wavelengths of light. This separation is large 
enough to ensure that ions can be 
addressed individually by a laser. The idea of a linear ion trap is the same 
as that of a Paul trap \cite{Paul90} which is already being used to trap single 
ions for very long times. Linear ion traps are already working and it is possible 
to trap strings of $30$ ions or more in them \cite{Walther92}. However, no 
linear ion trap quantum computer with $30$ ions has been realized yet.

What is the reason for that. 
The practical problem with the implementation of the linear ion trap quantum 
computer is the mechanical
degree of freedom of the ions. Although the ions are trapped along the 
axis of the linear ion trap they are not at rest, but oscillate around 
their equilibrium position. After having trapped the ions, the next step 
is then to cool them using methods of laser cooling which have recently 
been recognized by the Nobel Prize for physics \cite{Adams97,Itano95}. While 
it is fairly standard today to cool ions to temperatures of the order of millikelvin, 
it is very difficult to cool them to the necessary ground state of motion, 
i.e. to a state in which only the unavoidable motion due to the quantum 
mechanical uncertainty principle is present. A single ion has been 
cooled to its ground state of motion \cite{Wineland95} some years ago, 
while two ions have been cooled to the ground state of motion while
we were writing this review \cite{Wineland98}. Experimentalists 
in this field are optimistic that 
it will soon be possible to cool a string consisting of a few ions to 
the ground state of motion. 

As it is so difficult to cool the ions to the ground state of motion,
we need to have a good reason to try it. To see why we want to cool the 
ions to the ground state of motion 
one should remember that we want to implement quantum gates between 
different qubits. In the ion trap, these qubits are localized and we 
cannot really move them from one place to another. If we want to implement 
a quantum gate between two ions that are separated, e.g. one at the 
beginning of the string and one at the end, then we need some 'medium' 
that can be used for communication between these ions. Note that this 
communication is not classical but has to be quantum mechanical in 
nature as we want to establish quantum mechanical coherence between 
different ions. This communication is achieved by using the center-of-mass 
mode of the ions. If we excite the center-of-mass mode of the ions then all 
of the ions will oscillate in phase and therefore all of them will feel this 
oscillation simultaneously. This behaviour is illustrated in Fig. \ref{com}a. 
If the first ion is excited by a laser and the absorbed photon excites the
center-of-mass mode then even the ion at the end of the chain will feel this. Therefore even distant ions can 'communicate'. In the next section we will
explain how we can use the center-of-mass mode to generate a CNOT gate.

\subsection{The implementation of the CNOT gate}
This idea of using the center-of-mass mode as a 'bus' is the key 
ingredient in the ion trap quantum computer. It allows the implementation 
of two-bit gates such as a CNOT gate, for example. In the following we 
will explain how one can implement a CNOT gate in a linear ion trap 
computer. More complicated gates can be constructed, but as we pointed 
out in Section II a CNOT gate together with single-qubit rotations are 
sufficient to implement any unitary operation between quantum bits. 
Obviously single-qubit gates do not require the center-of-mass mode 
as a bus as they are implemented by manipulating a single ion with a 
suitably made laser pulse. In a CNOT gate it is essential that the 
two qubits (i.e.ions) interact and this is achieved by exciting the 
center-of-mass mode of all the ions in the linear ion trap.
Therefore, before we describe how to implement a full CNOT gate, we 
explain briefly how we can excite the center-of-mass mode of the ions
with a standing wave laser field. Let us first have a look at the 
energy levels of a single qubit and the center of mass mode which 
are given in Fig. \ref{ladder}. The vertical axis represents 
the energy of the joint system of ion and center-of-mass mode. 
The two levels $\{|00\rangle,|10\rangle \}$ on the far left are 
those with no phonons (i.e. excitations) in the center-of-mass mode. 
The lower state of the qubit is at energy zero, the upper state has 
energy $\hbar\omega$ where $\omega$ is the transition frequency between the
qubit levels. The next two energy levels $\{|01\rangle,|11\rangle \}$
represent the energies of the qubit when one phonon has been excited 
in the center-of-mass mode. The energy required to excite the 
center-of-mass mode is $\hbar\nu$ and this is usually a very small 
energy compared to the energy required to excite the qubit. $\nu$ is 
of the order of MHz as compared to the transition frequency in a qubit
which is of the order of $10^{15} Hz$. Before we give the Hamiltonian 
that describes the interaction between laser, ion and center-of-mass mode,
we have a qualitative look at the dynamics of the system. Imagine that an 
ion is localized in a standing wave laser field. This laser drives the qubit 
transition in the ion. If the laser has a frequency $\omega$ 
(shown by a vertical blue arrow in Fig. \ref{ladder}) then the laser will  
most likely induce transitions between the lower and upper state of the 
qubit without affecting the center of mass mode. This is simply because all 
other transitions, e.g. the $|00\rangle \leftrightarrow |01\rangle$ 
transition, are out of resonance. If, however, the 
frequency of the laser is $\omega-\nu$ (shown by a red arrow in Fig \ref{ladder}), 
then it predominantly generates transitions between the upper 
state of the ion and the vibrational state with $n$ excitations 
in the center-of-mass mode (state $|1 n\rangle$) and the ground 
state of the ion and the vibrational state with $n+1$ excitations 
in the center-of-mass mode (state $|0 n+1\rangle$). If both ion and 
center-of-mass mode are in the ground state, nothing at all happens. 
In summary we can see two effects. Firstly, a red detuned laser can change
simultaneously the electronic state of the qubit and the state of 
the center-of-mass mode. Secondly the dynamics can be conditional on the 
internal state of the ion, i.e. the qubit. If the qubit is in the lower 
state then there is no dynamics, if not then 
the laser induces transitions. One can easily see that this would
not be possible if the ions were not cooled to the ground state of 
the motion. If with high probability there is at least one phonon 
in the center-of-mass mode then the red detuned laser would always 
affect ions that are in the ground state.
  
This qualitative discussion neglects the importance of the position 
of the ion in the standing wave laser field. This is a very important 
factor, as it can be shown that an ion localized at the anti-node of 
the standing wave will, in leading order, interact with the laser 
without changing the excitation of the center-of-mass mode. If, on 
the other hand, the ion is localized at the node of the standing 
wave then in leading order both the internal degrees of freedom of 
the ion as well as the excitation of the center-of-mass mode are 
changed simultaneously. Qualitatively this can be understood in 
the following way. If the ion is at the anti-node of the field
then it does not see any photons. Therefore in order to interact 
with the field it has to change position and therefore it either 
has to absorb or emit a photon. Hence a change in its internal 
degree of freedom always requires a change in the motional degree 
of freedom of the ion. If the ion is localized at the node of the 
field then it is not necessary for it to move in order to see photons.
This qualitative reasoning can be corroborated by a precise derivation 
of the interaction Hamiltonian between ion, laser and center-of-mass 
mode. However, we refer the reader to the literature for this 
derivation \cite{Cohen92} and only state the results here.

First we assume that the ion is localized at the node of the standing 
wave laser field of frequency $\omega-\nu$. Then in leading order the 
Hamiltonian is given by
\begin{equation}
	H = \frac{\eta}{\sqrt{N}}\,\frac{\Omega}{2}
	\left[ |1\rangle\langle 0| a e^{i\phi} + 
	|0\rangle\langle 1|a^{\dagger} e^{-i\phi}\right]\; ,
	\label{4.1}
\end{equation}
where $N$ is the number of ions that are in the linear ion trap, 
$\Omega$ is the Rabi frequency of the laser, $\phi$ is the phase 
of the laser. The Lamb-Dicke parameter 
$\eta=(2\pi/\lambda)\sqrt{\hbar/2M\nu}$ describes how well the ions 
are localized and $a^{\dagger},a$ are creation and annihilation 
operator for excitations in the center-of-mass mode. 
This Hamiltonian is an approximation and represents only the 
first term in an expansion of the true Hamiltonian in terms of 
$\eta$. In addition, it neglects the interaction with 
other modes than the center-of-mass mode such as the one shown
in Fig. \ref{com}b. These are good approximations as $\eta$ is 
much smaller than unity and because other modes of oscillation 
have resonance frequencies different from $\nu$ and are therefore 
out of resonance with the laser.

If the ion is localized at the anti-node of the standing wave 
laser field of frequency $\omega$ then the Hamiltonian is given by
\begin{equation}
	H = \frac{\Omega}{2}
	\left[ |1\rangle\langle 0| e^{i\phi} + 
	|0\rangle\langle 1| e^{-i\phi}\right]\; .
	\label{4.2}
\end{equation}
The motional degree of freedom of the ions remains (in leading order) 
unchanged during the interaction with the laser. Again there will be 
higher order corrections in $\eta$ and off-resonant coupling terms 
to modes other than the center-of-mass mode.

Now let us see how we can implement a CNOT gate \cite{Cirac95}. For 
simplicity we assume that we have only two ions in the linear ion trap, 
as the whole procedure generalizes easily to more ions. We split the 
procedure into two parts, as it
then becomes more transparent. First we show how one can implement a 
controlled phase gate, i.e. a gate that flips the phase of the upper
state of the target qubit only if the control qubit is in the upper state.
Then we show that one can easily obtain a CNOT gate from this.

It is important that initially the center-of-mass mode is in the 
ground state. First we place the control qubit at the node of a
standing wave laser filed. The interaction is described by the 
Hamiltonian Eq. (\ref{4.1}). We apply the laser pulse on the control 
bit for a time $t=\pi/(\Omega\eta/\sqrt{2})$ so that it produces a 
$\pi$-pulse. 
If in $|xyz\rangle$, $x$ describes the state of the control qubit,
$y$ the state of the target qubit and $z$ the state of the center-of-mass
mode then we find that the laser pulse has the following effect. 
\begin{eqnarray}
	|00\rangle |0\rangle &\rightarrow& |00\rangle|0\rangle\\ \nonumber
	|01\rangle |0\rangle &\rightarrow& |01\rangle|0\rangle\\ \nonumber
	|10\rangle |0\rangle &\rightarrow& -i |00\rangle|1\rangle\\ \nonumber
	|11\rangle |0\rangle &\rightarrow& -i |01\rangle|1\rangle \;\; .
\end{eqnarray}
Note that the action of the laser depends on the state 
of the control ion. If the control qubit is in the ground state then 
nothing happens, but if it is in the upper state then it goes to the 
ground state and simultaneously the center-of-mass mode is excited.

Now we manipulate the target qubit which we place at the node of a
standing wave laser field. In this step we make use of the fact 
that in a real ion we have many atomic levels available. In Fig. 13 we
see the qubit levels $0$ and $1$ and an additional energy level $2$.
We now couple the lower level $0$ of the qubit to the auxiliary level $2$
again using a Hamiltonian of the form Eq. (\ref{4.1}) (only that we replace 
$1$ by $2$). We apply the laser for a time $t=2\pi/(\Omega\eta/\sqrt{2})$ 
so that we effectively perform a full $2\pi$ rotation. This means that
no population at all ends up in level $2$. The effect of this transition
is that the ground state of the target bit flips its phase  
if the center-of-mass mode is excited. Therefore we obtain
\begin{eqnarray}
	   |00\rangle |0\rangle &\rightarrow&    |00\rangle|0\rangle\\ \nonumber
	   |01\rangle |0\rangle &\rightarrow&    |01\rangle|0\rangle\\ \nonumber
	-i |00\rangle |1\rangle &\rightarrow&  i |00\rangle|1\rangle\\ \nonumber
	-i |01\rangle |1\rangle &\rightarrow& -i |01\rangle|1\rangle \;\; .
\end{eqnarray}
In the next step we apply again a pulse of duration 
$t=\pi/(\Omega\eta/\sqrt{2})$ to the control qubit using 
Hamiltonian Eq. (\ref{4.1}). This results in the total transformation
\begin{eqnarray}
	|00\rangle |0\rangle &\rightarrow&  |00\rangle|0\rangle\\ \nonumber
	|01\rangle |0\rangle &\rightarrow&  |01\rangle|0\rangle\\ \nonumber
	|10\rangle |0\rangle &\rightarrow&  |10\rangle|0\rangle\\ \nonumber
	|11\rangle |0\rangle &\rightarrow& -|11\rangle|0\rangle \;\; .
	\label{tran}
\end{eqnarray}
Therefore we have implemented a controlled phase gate.
Now let us see why this transformation is really the basic building 
block for a CNOT gate. Consider the different set of basis states
\begin{equation}
	|\pm\rangle = (|0\rangle \pm |1\rangle )/\sqrt{2} \;\; .
	\label{basis}
\end{equation}
If we rewrite the state of the target qubit in the basis Eq. (\ref{basis}) then
the transformation Eq. (\ref{tran}) reads as
\begin{eqnarray}
	|0+\rangle |0\rangle &\rightarrow&  |0+\rangle|0\rangle\\ \nonumber
	|0-\rangle |0\rangle &\rightarrow&  |0-\rangle|0\rangle\\ \nonumber
	|1+\rangle |0\rangle &\rightarrow&  |1-\rangle|0\rangle\\ \nonumber
	|1-\rangle |0\rangle &\rightarrow&  |1+\rangle|0\rangle \;\; .
\end{eqnarray}
This is the action of a CNOT gate. So, all we need to do is to rotate the 
$\{|0\rangle,|1\rangle \}$ basis of the target bit into the 
$\{|+\rangle,|-\rangle \}$ basis, then perform the controlled phase gate, 
and finally we rotate the basis back to $\{|0\rangle,|1\rangle \}$. This 
generates a CNOT gate in the $\{|0\rangle,|1\rangle \}$ basis. 
The rotation  between the basis sets can be achieved using the 
Hamiltonian Eq. (\ref{4.2}), i.e. with a standing wave that has the target 
ion at its anti-node. For this we just have to chose the phase of the laser 
to be $\phi=-\pi/2$ and perform a $\pi/2$ pulse, i.e. a pulse with the length
$t=\pi/(2\,\Omega)$. Going back from the $\{|+\rangle,|-\rangle \}$ basis to 
the $\{|0\rangle,|1\rangle \}$ basis is then done by a $-\pi/2=3\pi/2$ pulse.
Therefore we are able to generate a CNOT gate as well as single qubit 
gates and this is all we need to implement any unitary transformation
between qubits. 

In our analysis we have made quite a number of simplifying assumptions 
some of which we have already mentioned. The Hamilton operators Eqs.
(\ref{4.1},\ref{4.2}) are only the leading order terms in an expansion 
with respect to the Lamb-Dicke parameter $\eta$.
In addition we have only taken into account the center-of-mass mode although
there are many other modes that might also get excited. 
We also neglected any spontaneous emission from the ions and losses 
of excitations of the center-of-mass mode.
That these are reasonable approximations can be seen most convincingly from
the fact that a CNOT gate has been realized 
experimentally in an ion trap, albeit by a different scheme using only one ion 
\cite{Monroe95}. In this realization the main source of noise was technical
noise.

If we want to perform many quantum gates on many quantum bits then,
however, the effect of errors will become much more serious. Therefore
it is important to analyze the effect of noise on a quantum computer 
and to find ways to circumvent or
correct them. This problem will be addressed in the next section.

\section{Decoherence and Quantum Computation}

In the last section we saw that in principle, we can implement a quantum 
computer in a linear ion trap. A single quantum gate has been demonstrated 
in this scheme \cite{Monroe95} and it is expected that soon a few quantum 
gate operations will be performed on a few qubits. However, scaling up this
implementation (i.e. implementing large numbers of quantum gates on many 
qubits) is not easy because noise from all kind of sources will disturb 
the quantum computer. Of course also a classical computer suffers from 
the interaction with a noisy environment and nevertheless works very well.
The advantage of a classical computer is, however, that it is a classical
device in the sense that one bit of information is represented by the 
presence or absence of a large number of electrons ($10^8$ electrons 
for example). Therefore a small fluctuation in the number of electrons 
does not disturb the computer at all. On the contrary, in a quantum 
computer the qubit is stored in the electronic degree of freedom of a 
single atom. Even worse than that, a quantum computer crucially depends 
on the survival of quantum mechanical superposition states which are 
notoriously sensitive to decoherence and dissipation \cite{Zurek91}. 
This makes a quantum computer extremely sensitive to small perturbations 
from the environment. It has been shown that even rare spontaneous 
emissions from a metastable state rule out long calculations unless new 
ideas are developed \cite{Plenio96,Plenio97a,Plenio97b}.
However, from classical information theory we know that it is possible
to correct for errors by introducing redundancy \cite{SW48}.
Although initially there were doubts as to whether it would be possible 
to generalize these ideas to quantum mechanics, quantum error correction 
codes have been developed recently. 

In this section we will give an example to illustrate how a quantum 
mechanical superposition state is very sensitive to noise using a special 
example of a noise process. Then we will discuss a general method that 
can be used to understand noise and its origin at the quantum level. 
Finally we explain the ideas of quantum error correction that have been 
developed to correct for the detrimental influence of noise in quantum 
computation. 

\subsection{Decoherence of entangled states}

To illustrate that entangled states are particularly sensitive to noise
we assume that each two-level system interacts independently with the 
environment which is usually a very good assumption. The source of 
errors are so-called phase errors, represented by the operator
\begin{equation}
	\sigma_z = \left( \begin{array}{lr} 1 & 0 \\ 0 & -1 \end{array} \right) \;\; .
	\label{5a2}
\end{equation}
This means that if such an error occurs the phase of the ground state 
remains unchanged while the excited state flips its phase, i.e. 
$\sigma_z |0\rangle = |0\rangle$ and $\sigma_z |1\rangle = -|1\rangle$. 
Note that $\sigma_z$ produces a discrete phase change. In reality 
we can find any arbitrary phase change given by the operator
\begin{equation}
	\sigma_{\phi} = \left( \begin{array}{lr} 1 & 0 \\ 0 & e^{i \phi}
			                         \end{array} \right) \;\; .
	\label{5a2a}
\end{equation}
However, the operator $\sigma_{\phi}$
corresponding to this phase change can be reexpressed as a linear combination
of the identity operator and the operator $\sigma_z$. Therefore the analysis
using only $\sigma_z$ is sufficient. 
A physical source of phase-noise can be that the ion that stores the 
qubit collides with atoms from the background gas. During this event 
usually no energy is exchanged but nevertheless the relative phase of 
states can be changed. We assume that the rate at which a single 
two-level system collides with atoms from the background gas is $\gamma$.
If we prepare the qubit in a superposition state
\begin{equation}
	|\psi\rangle = \frac{1}{\sqrt{2}}(|0\rangle + |1\rangle)
\end{equation} 
then the relative phase will be randomized at a rate $\gamma$. After a 
time $t$ the state will have evolved into a statistical mixture
\begin{equation}
	\rho = \left( \begin{array}{lr} 
		\frac{1}{2} & \frac{1}{2} e^{-\gamma t} \\  
                \frac{1}{2} e^{-\gamma t} &  \frac{1}{2} \end{array} \right)\; .
\end{equation}
Now let us consider $n$ qubits. We assume that they are prepared in the state 
\begin{equation}
	|\psi\rangle = (|\underbrace{0\ldots 0}_{n\, times}\rangle + 
		      |\underbrace{1\ldots 1}_{n\, times}\rangle)/\sqrt{2} \;\; .
	\label{5a3}
\end{equation}
In our error model, we observe that the relative phase of this highly 
entangled state is randomized at a rate that is much higher than $\gamma$. 
After a time $t$ the state is 
\begin{equation}
	\rho = \left( \begin{array}{lr} 
		 \frac{1}{2} & \frac{1}{2} e^{-n \gamma t} \\  
                 \frac{1}{2} e^{-n\gamma t} &  \frac{1}{2} \end{array} \right)\; .
\end{equation}
when it is written in the basis $\{|0\dots 0\rangle,|1\dots 1\rangle \}$ 
basis. This accelerated randomization of the relative phase derives from 
the fact that an error in any one of the qubits switches the phase of the 
state Eq. (\ref{5a3}). Therefore we have $n$ possibilities to destroy the 
phase which results in an effective decoherence rate of $n \gamma$. 

The state Eq. (\ref{5a3}) is a particular example and not all states show 
the same sensitivity to decoherence. Nevertheless this example should make
clear that a quantum computer is very sensitive to noise and, even worse, 
it becomes more and more sensitive to noise the larger the quantum computer
becomes.

\subsection{A general error model}

After this specific example of an error process, we will now consider a 
more general method to describe errors in qubits. In this description we 
take the state of the environment explicitly into account. In the example 
given above, that would mean that we keep track of the state of the 
background gas atom that has collided with our qubit. This is extremely 
difficult in practice but there is no physical principle that forbids it. 
In the following we will see that after an interaction with the environment
the qubit becomes entangled with the environment. It is the fact that we 
do not have any knowledge about the environment that destroys the coherence 
of our quantum state. This description will also pave the way to the 
understanding of methods that are able to correct errors that are generated 
in a quantum states.

We assume that the initial state $|\psi\rangle$ of our qubit is pure and 
that the environment is in the pure state $|e\rangle$. Initially 
the qubit and the environment are completely uncorrelated, i.e. the total 
state of the combined system is
\begin{equation}
	|\psi_{tot}\rangle = |\psi\rangle \otimes |e\rangle \;\; .
	\label{5a4a}
\end{equation}
Now, however, the qubit will interact with the environment. There are many
possibilities, such as exchanging energy in the form of photons or the 
above mentioned phase errors due to collisions etc. In general the 
interaction with the environment is simply a joint unitary transformation 
between qubit and environment. Such a unitary transformation can be written 
in many different ways. A particularly useful way is via the Pauli spin 
operators which are of the form
\begin{equation}
	\sigma_x = \left( \begin{array}{cc} 0 & 1 \\  1 &  0 \end{array} \right)\;
	\sigma_y = \left( \begin{array}{cc} 0 & i \\ -i &  0 \end{array} \right)\;
	\sigma_z = \left( \begin{array}{cc} 1 & 0 \\  0 & -1 \end{array} \right)\;
	\sigma_0 = \left( \begin{array}{cc} 1 & 0 \\  0 &  1 \end{array} \right)\; 
	\label{5a5}
\end{equation}
After some time the total state of system and environment is then given by \cite{Chiara96}
\begin{equation}
	|\psi_{tot}\rangle = \sigma_x|\psi\rangle {\cal T}_{x}|e\rangle 
                           + \sigma_y|\psi\rangle {\cal T}_{y}|e\rangle
                           + \sigma_z|\psi\rangle {\cal T}_{z}|e\rangle
                           + \sigma_0|\psi\rangle {\cal T}_{0}|e\rangle \;\; ,
	\label{5a6}
\end{equation}
where ${\cal T}_{ij}$ stand for a unitary transformation acting on the 
state of the environment $|e\rangle$ only. Note that we have four 
different errors represented by the operators $\sigma_i$. 
This stands in contrast to the classical case where we have only one 
sort of error, namely the bit flip $0 \leftrightarrow 1$. Every possible 
error $\sigma_i$ is correlated to a specific state of the environment. So 
far we still have a pure state for the entire system including the 
environment and in principle it would be possible to recover the initial 
state of the qubit. Now, however, our ignorance of the state of the 
environment comes into play. Obviously, if we have no information about 
the state of the environment (which is usually the case) then we cannot 
find out what error has occurred and therefore we are unable to correct 
the error in the quantum state. 

Now the question arises whether it is possible to correct errors that 
have occurred when we do not have access to the states of the environment! 
Is this possible at all? This question will be answered positively in the 
next subsection.

\subsection{Quantum error correction} 
In the previous subsection we have raised the question of whether we 
can correct errors in qubits as we can in classical communication and 
computation. Initially there were doubts that errors in quantum mechanical 
states can be corrected, but soon it become clear that this is in fact 
possible. The first quantum error correction code was discovered in 
1995 by Shor \cite{Shor95} and later a more general theory of quantum 
error correction was developed
\cite{Calderbank96,Chiara96,Knill97,Steane96a,Steane96b}. This development 
has continued and has led to an avalanche of different codes that 
were optimized in different respects and adapted to special situations 
\cite{Braunstein97,Laflamme96,Lloyd97,Plenio97c,Vaidman96}. New mathematical
techniques have been developed which are particularly suited to the study
of quantum error correction codes
\cite{Bennett96,Calderbank96,Calderbank96b,Calderbank97,Gottesman96,Gottesman97}
\cite{Gottesman97a,Gottesman97b,Knill97,Rains96a,Rains96b,Rains97a,Rains97b}
\cite{Rains97c,Rains97d,Shor96b,Steane96c}.

How does quantum error correction work? Let us reconsider the example
in subsection B where a qubit has been interacting with an environment.
This interaction led to a correlation of the qubit with the environment,
e.g. if the qubit collides with a background gas atom, their internal states
become correlated. In principle such an error could be corrected if we could 
obtain complete knowledge of the joint state of the two systems. The fact 
that really makes it impossible to correct the errors
is that we usually have no access to the information that is stored in the 
environment, e.g. the photon that has been emitted has not been detected by us.
We do not know which error has occurred and therefore we are unable to correct it. 
So, we see that the source of irreversibility is the combination of entanglement 
of the qubit with the environment and the loss of information about the state of
the environment. 

How can we combat this loss of information? We could for example entangle 
our information-carrying qubit with some auxiliary qubits such that we 
can distribute the information about the information-qubit over many 
auxiliary qubits. To understand why this might work imagine the information 
qubit interacts with the environment and we have no access to the environment.
Of course, the state of the qubit is now perturbed, but luckily this time 
we have some additional information about its original state in our 
auxiliary qubits that we can, unlike the environment, still access. The 
idea is now to correct the state of our qubit using this additional 
information. This is, in a very rough form, the idea of quantum error
correction. 

Now we present a simple example of a quantum error correction code 
that protects one qubit against the occurrence of a single amplitude error, 
i.e. the $0 \leftrightarrow 1$ bit flip. Obviously, as just pointed out, we 
need additional qubits so that we can distribute over many qubits the 
information that was previously in the state of one qubit. In this case 
two additional qubits are necessary and sufficient. This amplitude error 
correcting code is presented in Fig. \ref{ampl} where we have given the 
quantum network necessary to implement the quantum code and the subsequent 
error correction. This code has the property that it encodes the states 
$|0\rangle$ into state $|000\rangle$ and $|1\rangle$ into state $|111\rangle$. 
A superposition $\alpha|0\rangle + \beta |1\rangle$ is therefore encoded 
as $\alpha|000\rangle + \beta |111\rangle$. When the first bit is in the
state $|\psi\rangle$ and the second and third qubit are in the state 
$|00\rangle$ then it is easily checked that this encoding is performed 
by the network on the left hand side in Fig. \ref{ampl}. After the encoding, 
one waits, and an error may occur during that time that is indicated by 
the box in Fig. \ref{ampl}. Subsequently we decode our state using two
CNOT gates and then the error correction is performed in the 
last step using a Toffoli gate. Let us check whether the 
proposed method really works. Obviously when there has been no error
at all then at the end the first qubit is recreated in the right state.
What happens if the first bit suffers an amplitude error, i.e. it suffers 
the bit flip $0\leftrightarrow 1$? It is easy to see that before the 
application of the Toffoli gate the state of the three qubits will be
$(\alpha|1\rangle + \beta|0\rangle)\otimes |11\rangle$. The second and 
third bit indicate that an error has occurred in the first qubit and the 
subsequent Toffoli gate corrects this error by flipping the first qubit. 
We can check that for errors in the second or third qubit the state just 
before the application of the Toffoli gate will be 
$(\alpha|0\rangle + \beta|1\rangle)\otimes |10\rangle$ 
or $(\alpha|0\rangle + \beta|1\rangle)\otimes |01\rangle$. Therefore the 
Toffoli gate will leave the first bit unaffected. In conclusion we see 
that a single amplitude error can be corrected using the additional 
information about the environment stored in the auxiliary qubits. Note 
that we did not need to measure the state of the second and third qubit 
-- all our encoding, decoding and error correction is performed by unitary
transformations. Nevertheless at some point irreversibility catches us. 
Indeed if we want to reuse the two auxiliary qubits for the next round 
of quantum error correction then we have to prepare them in a standard 
state, e.g. the state $|00\rangle$. That means that we either have to
dissipate irreversibly their energy to the environment, or we have to 
measure their state and then change their state by a unitary transformation.

On the other hand we could have done the error correction without the 
Toffoli gate. Instead we could perform a measurement of the state of 
the auxiliary quantum bits in the 
$\{|0\rangle,|1\rangle\}$ basis. If we find the state $|11\rangle$ then
we invert the first bit, otherwise we leave it unaffected.
Finally have to reset the auxiliary bits to the ground state using a 
unitary transformation. 

We have seen how to implement an amplitude error 
correcting code. But amplitude errors are not the only possible errors 
in quantum mechanics as we have already seen above -- phase errors are 
a different possibility. Does the above code work when we have a phase
error? A quick check reveals that the network given in Fig. \ref{ampl} 
is not able to correct phase errors at all (Try it!). But one can easily 
change the network to allow the correction of phase errors. One just has 
to observe that the following relation holds
\begin{equation}
	\sigma_z = \left( \begin{array}{cc} 0 & 1 \\  1 &  0 \end{array} \right)
                 = \frac{1}{2}
                   \left( \begin{array}{cc} 1 & 1 \\  1 &  -1 \end{array} \right)\cdot
		   \left( \begin{array}{cc} 0 & 1 \\  1 &  0 \end{array} \right)\cdot
                   \left( \begin{array}{cc} 1 & 1 \\  1 &  -1 \end{array} \right)
\end{equation}
This means that if we consider a phase error ($\sigma_z$ operator) 
in a rotated basis then it appears as a amplitude error and vice versa. 
This new basis that we obtain from the $\{|0\rangle,|1\rangle\}$ by using 
a Hadamard transformation is given by
$|\tilde{0}\rangle = (|0\rangle + |1\rangle)/\sqrt{2}$ and 
$|\tilde{1}\rangle = (|0\rangle - |1\rangle)/\sqrt{2}$.
In this new basis a phase error has the effect of an amplitude error, i.e. 
it has the effect $\sigma_z|\tilde{0}\rangle= |\tilde{1}\rangle, 
\sigma_z|\tilde{1}\rangle= |\tilde{0}\rangle$.
Therefore instead of 
encoding the state $\alpha |0\rangle + \beta|1\rangle$ as shown 
above we encode it as 
\begin{equation}
	\alpha |0\rangle + \beta|1\rangle \rightarrow 
	\alpha |\tilde{0}\tilde{0}\tilde{0}\rangle + \beta|\tilde{1}
\tilde{1}\tilde{1}\rangle \; \; .
\end{equation} 
Therefore we obtain an phase error correcting code using a network 
as shown in Fig \ref{phase}. This construction of the phase error 
correcting code from the amplitude error correcting code reveals 
an important principle that can be used to generate quantum error 
correcting codes that are able to correct one general error, i.e. 
one phase errors ($\sigma_z$), one amplitude errors ($\sigma_x$) or 
a combination of both which is represented by the operator $\sigma_y$. 
If we want to 
achieve this, we need a code that, looked at in the 
$\{|0\rangle,|1\rangle\}$ basis, corrects amplitude errors and 
when written in the $\{|\tilde{0}\rangle,|\tilde{1}\rangle\}$ 
basis is also an amplitude error correcting code. Then the basic 
idea is first to check in the unrotated basis whether there has 
been an amplitude error. Then one rotates the state into the basis
$\{|\tilde{0}\rangle,|\tilde{1}\rangle\}$ and again one checks 
whether in this basis we can find an amplitude error. Codes that 
have these properties can be found (although you need some knowledge 
of classical error correction). This approach can be found in much greater
detail in \cite{Steane96a,Steane96b}. This approach is attractive, as it 
can be shown that one can construct quantum error correcting codes 
from classical error correcting codes. These ideas can be generalized 
to incorporate the most general quantum error correcting codes but for 
details of the mathematical principles behind this construction
and actual examples we refer the interested reader to the literature \cite{Steane96a,Steane96b,Gottesman97b}. It should also be noted 
that experimental progress has recently been made. Using a nuclear
magnetic resonance implementation of a quantum computer the 
simple three bit code presented in subsection VI.C has been 
demonstrated recently \cite{Cory98}.
 
So far we have seen that it is possible to use quantum error correction 
to correct errors that have occurred in a quantum bit. Errors will of 
course also appear during the operation of quantum gates. Therefore 
the question arises of
what happens when errors appear during one of the quantum gates that 
are performed to generate the error syndrome and the subsequent error
correction. Such an error can make the error correction fail and it would
be important to establish whether it is possible to perform quantum 
error correction in a 'fault-tolerant' form. Note also that an error during the
operation of a two-bit gate may actually result in two errors. This can easily
be seen at the example of a CNOT-gate. If the control-qubit suffers an
amplitude error before the CNOT gate is performed then after the CNOT gate
both the control and the target bit have suffered an amplitude error,
(see Fig. \ref{error}) and check. Therefore errors during gate operations
are particularly destructive and obviously methods have to be developed
that avoids this multiplication of errors. 
In fact, such methods have been developed. However, they are quite involved 
and it would be beyond the scope of this article to review them. The 
interested reader is referred for details to the literature on 
fault-tolerant quantum computation
\cite{Aharonov96,Gottesman97a,Knight97,Shor96,Preskill97,Steane97} and 
its special case of fault-tolerant quantum error correction 
\cite{DiVincenzo96}.

\section{Summary and Future Prospects}

In this review we have seen that the laws of 
quantum mechanics imply a different kind
of information processing to the traditional 
one based on the laws of classical physics.
The central difference, as we emphasised, was in the
fact that quantum mechanics allows physical
systems to be in an entangled state, a phenomenon
non-existent in classical physics. This leads 
to a quantum computer being able to solve certain
tasks faster than its classical counterpart.
More importantly, factorization of natural 
numbers into primes can be performed efficiently
on a quantum computer using Shor's algorithm, whereas 
it is at present considered to be intractable
on a classical computer. However, to realize a quantum 
computer (or indeed any other computer) we have to
have a physical medium in which to store and
manipulate information. It is here that 
quantum information becomes very fragile and
it turns out that the task of its storage and 
manipulation requires a lot of experimental ingenuity.
We have presented a detailed account of the linear
ion trap, one of the more promising proposals
for a physically realizable quantum computer.
Here information is stored into electronic states of
ions, which are in turn confined to a
linear trap and cooled to their ground  
state of motion. Laser light is then used to manipulate
information in the form of different electronic
transitions. However, the uncontrollable interactions
of ions with their environment induce various
errors known as decoherence (such as e.g. spontaneous emission in ions)
and thus severely limit the power of computation.       
We have than shown that there is a method to
combat decoherence during computation known under the name of 
quantum error correction. This then leads to the notion
of fault tolerant quantum computation, which is a 
method of performing reliable quantum computation 
using unreliable basic components (e.g. gates) 
providing that the error
rate in this components is below a certain 
allowed limit. Much theoretical work has been
undertaken in this area at the moment and there is
now a good understanding of its powers and limitations.
The main task is now with the experimentalists to 
try to build the first fully functional quantum 
computer, although it should be noted that 
none of the present implementations appear
to allow long or large scale quantum computations and
a breakthrough in technology might be needed.

Despite the fact that at present large computational 
tasks seem to lie in the remote future, there is a lot
of interesting and fundamental physics that can be 
done almost immediately with the present technology.  
A number of practical information transfer protocols 
use methods of quantum computation. One example is
teleportation involving two parties usually referred
to as Alice and Bob. Initially Alice and Bob share an
entangled state of two qubits (each one having a single qubit)
Alice then receives a qubit in a 
certain (to her unknown) state which she wants 
to ``transmit" this state to Bob without actually sending the 
particle to him. She can do this by performing a simple
quantum computation on her side, and communicating
its result to Bob. Bob then performs the appropriate
quantum computation on his side, after which his qubit 
assumes the state of the Alice's qubit and the teleportation
is achieved (for details see \cite{tele}). It should 
be noted that this experiment has been performed recently
by a group in Innsbruck who achieved a successful 
teleportation of a singe qubit  
\cite{Zei98}. Again, since
entangled states are non-existent in classical physics
this kind of protocol is impossible, leading to another 
advantage of quantum information transfer over its
classical analogue. An extension of this idea leads 
to more than two users, say $N$, this time sharing entangled states
of $N$ qubits. If each of the users does a particular
quantum computation locally, and then they all communicate
their results to each other, then more can be achieved 
than if they did not share
any entanglement in the first place. This idea is known 
as distributed
quantum computation and is currently being developed 
\cite{Cirac98,Grover97}.  There
is a number of other interesting protocols and 
applications of quantum computation that have 
either been achieved or are within experimental 
reach , e.g. \cite{Bose98,Huelga97}. We hope that quantum factorization 
and other
large and important quantum computations will be
realized eventually. Fortunately, 
there is a vast amount of effort and ingenuity being applied to these 
problems, and the future possibility of a fully functioning 
quantum computer still 
remains very much alive. En route to this realization, we will discover a great deal of 
new physics involving entanglement, decoherence and the preservation 
of quantum superpositions.

\noindent
\section{ Acknowledgements}. This work was supported in part by the European TMR
Research Network ERB 4061PL95-1412, the European TMR Research Network 
ERBFMRXCT96066, a Feodor Lynen grant of the Alexander von Humboldt Stiftung,
the EPSRC and the Knight trust. The authors would like to thank Peter 
Knight for helpful comments on the manuscript.

\newpage

\begin{figure}
\vspace{4mm}
\epsfxsize12.cm
\centerline{\epsfbox{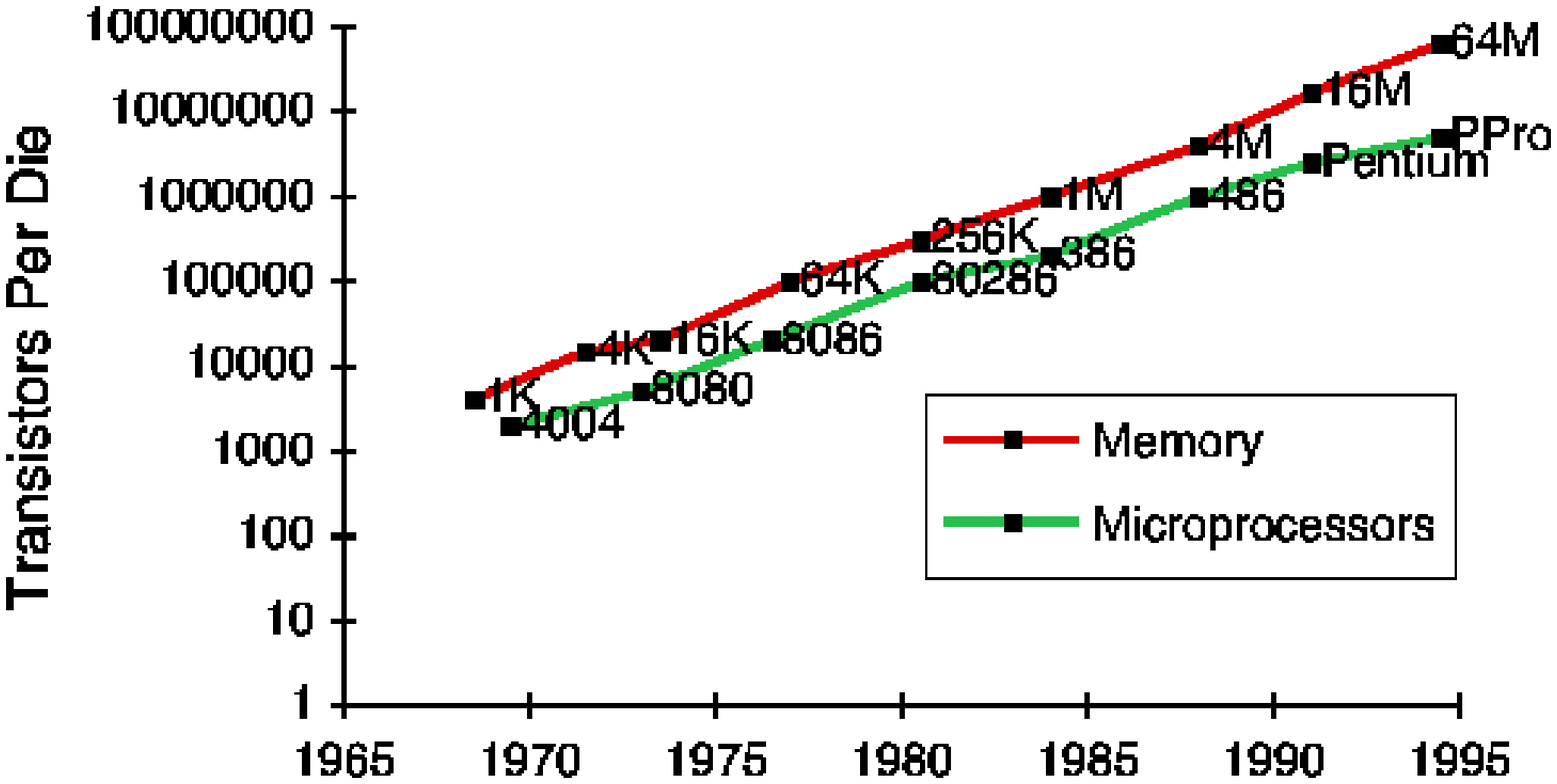}}
\vspace{2mm}
\caption[fo1b]{ \small 
Moore's Law: the number of transistors on a die or a micro--chip 
doubles every year}
\label{transistor}
\end{figure}

\begin{figure}
\vspace{4mm}
\epsfxsize10.cm
\centerline{\epsfbox{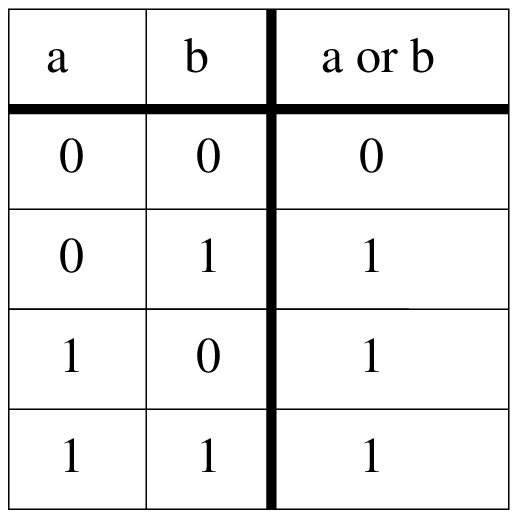}}
\vspace{2mm}
\caption[fo1b]{ \small 
The logical OR operation is a two input one output gate. This is
a typical example of a classical irreversible gate: from the knowledge
of the output ($a$ OR $b$) it is impossible to infer the input in general.}
\label{orgate}
\end{figure}

\begin{figure}
\vspace{4mm}
\epsfxsize10.cm
\centerline{\epsfbox{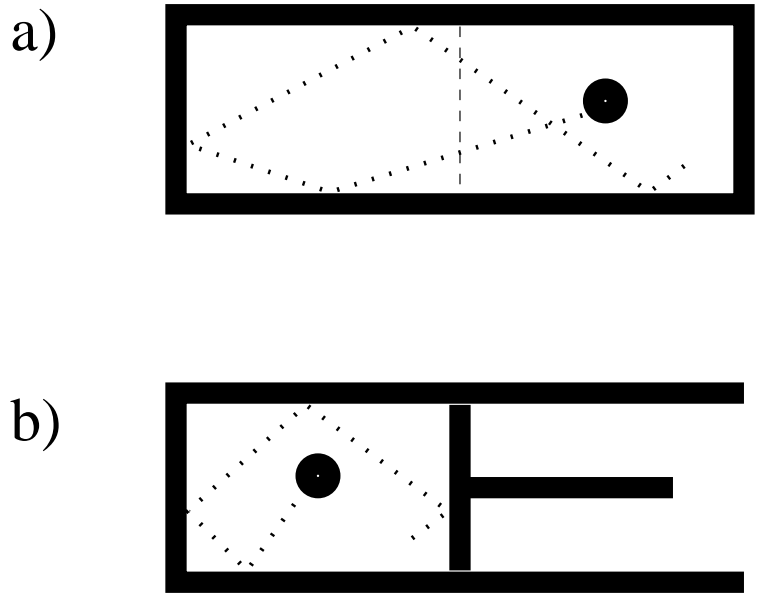}}
\vspace{2mm}
\caption[fo1b]{ \small 
  a) An atom-in-a-box representation of a bit: if the atom is in the
left hand half of the box this represents a logical 0, whereas if 
it is in the right hand half this represents a logical 1; b) confining
the atom to one side, by pushing the piston, resets the bit to 
one of the two values thereby dissipating $kT \ln 2$ of heat.}
\label{diss}
\end{figure}

\begin{figure}
\vspace{4mm}
\epsfxsize10.cm
\centerline{\epsfbox{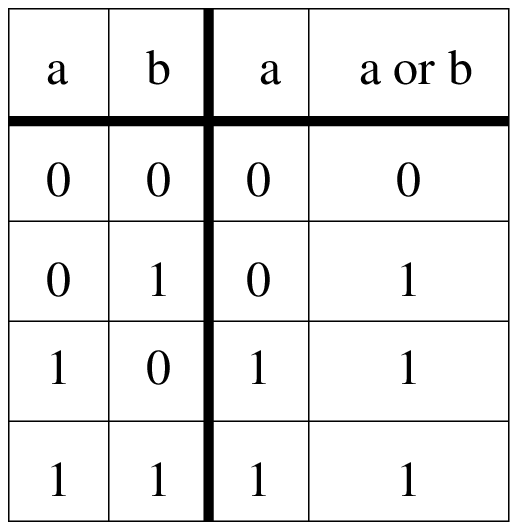}}
\vspace{2mm}
\caption[fo1b]{ \small 
We can make the OR gate reversible by preserving a part of the input,
in the above case the bit $a$.}
\label{revor}
\end{figure}

\newpage 

\begin{figure}
\vspace{4mm}
\epsfxsize10.cm
\centerline{\epsfbox{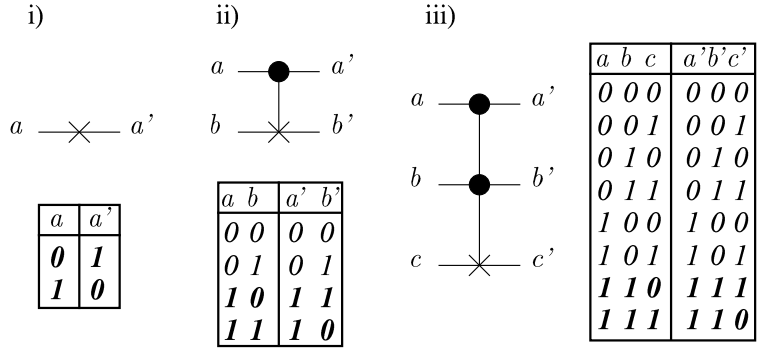}}
\vspace{2mm}
\caption[fo1]
{ \small Truth tables and graphical representations of the elementary
  quantum gates used for the construction of more complicated quantum
  networks. The control qubits are graphically represented by a dot,
  the target qubits by a cross. i) {\small\sf NOT} operation. ii)
  Control--{\small \sf NOT}.  This gate can be seen as a ``copy
  operation'' in the sense that a target qubit ($b$) initially in the
  state $0$ will be after the action of the gate in the same state as
  the control qubit. iii) Toffoli gate. This gate can also be seen as
  a Control--control--{\small \sf NOT}: the target bit ($c$) undergoes
  a {\small \sf NOT} operation only when the two controls ($a$ and
  $b$) are in state $1$. }
\label{basicgates}
\end{figure}

\begin{figure}
\vspace{4mm}
\epsfxsize12.cm
\centerline{\epsfbox{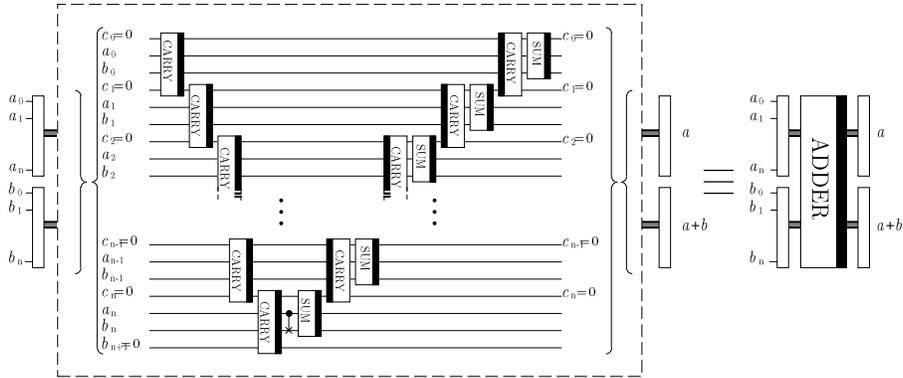}}
\vspace{2mm}
\caption[fo2]{ \small 
  Plain adder network. In a first step, all the carries are calculated
  until the last carry gives the most significant digit of the result.
  Then all these operations apart from the last one are undone in
  reverse order, and the sum of the digits is performed
  correspondingly. Note the position of a thick black bar on the right
  or left hand side of basic carry and sum networks.  A network with a
  bar on the left side represents the reversed sequence of elementary
  gates embedded in the same network with the bar on the right side.  }
\label{plainadder}
\end{figure}

\begin{figure}
\vspace{4mm}
\epsfxsize10.cm
\centerline{\epsfbox{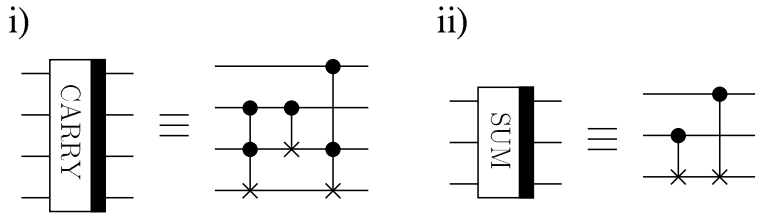}}
\vspace{2mm}
\caption[fo1b]{ \small 
  Basic carry and sum operations for the plain addition network. i)
  the carry operation (note that the carry operation perturbs the
  state of the qubit $b$). ii) the sum operation.}
\label{carrysum}
\end{figure}


\begin{figure}
\vspace{4mm}
\epsfxsize10.cm
\centerline{\epsfbox{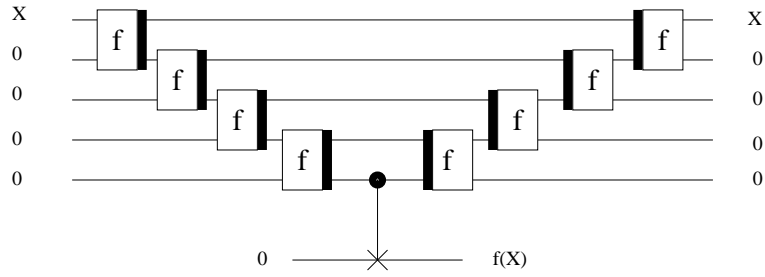}}
\vspace{2mm}
\caption[fo1b]{ \small 
  How to get rid of garbage bits once the computation is performed.
The network is the same in the classical and the quantum case, with
the exception that quantum bits can, of course, be in an entangled state. 
Thick black bars on the left represent the reverse operation, i.e.
the computation of $f^{-1}$}
\label{rubbish}
\end{figure}

\begin{figure}
\vspace{4mm}
\epsfxsize10.cm
\centerline{\epsfbox{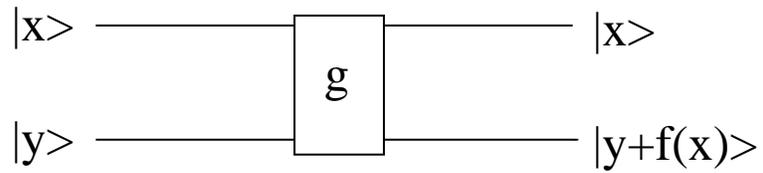}}
\vspace{2mm}
\caption[fo1b]{ \small 
  Quantum network for Deutsch's algorithm: the second qubit can be 
treated as a hardware of the quantum computer, so that there is, in fact,
only one input qubit $|x\rangle$. The function g leaves the first qubit
unchanged whereas the second one computes $y\oplus f(x)$}
\label{deutsch}
\end{figure}

\begin{figure}
\vspace*{1.cm}
\epsfxsize10.cm
\centerline{\epsfbox{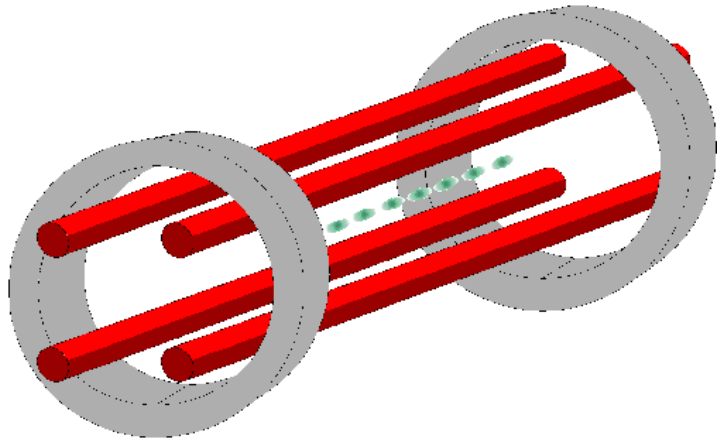}}
\vspace*{-3.cm}
\caption[fo1b]{Schematic picture of a linear ion trap computer. Electrodes generate a
time dependent electric field which generates an effective potential such that
a string of ions (the blue dots in the middle of the trap) is trapped. The 
motion of the ions, and in particular the center of mass mode, has to be cooled 
to its ground state. The center of mass mode then acts as a bus that allows
to generate interactions between any two ions.}
\label{lintrap}
\end{figure}

\begin{figure}
\vspace*{1.cm}
\centerline{\epsfbox{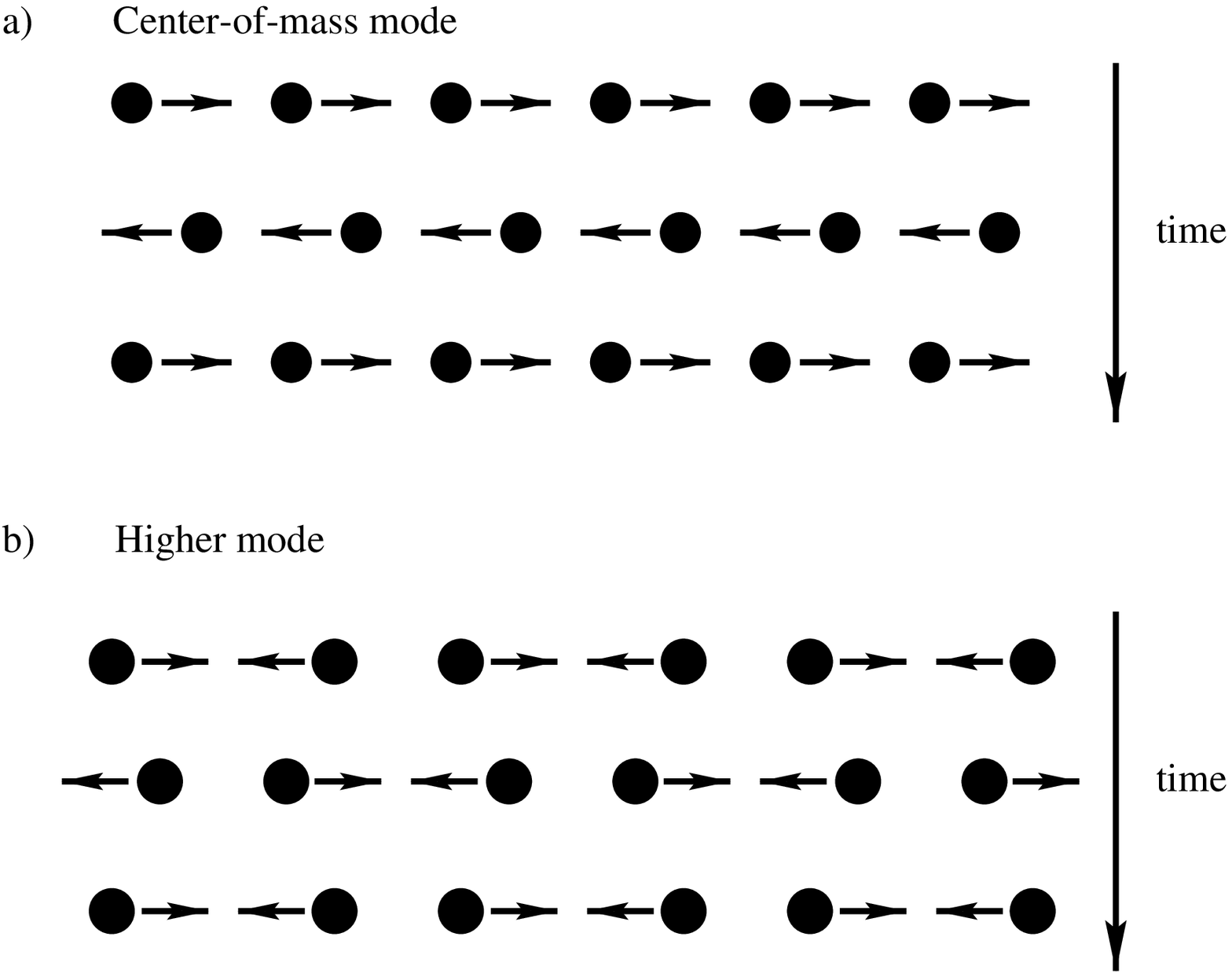}}
\caption[fo1b]{In part a) the center-of-mass mode is illustrated. All the ions oscillate
with the same phase. In part b) a mode of higher frequency is given. Here the ions
have different phases and their relative distances change.}
\label{com}
\end{figure}

\begin{figure}
\centerline{\epsfbox{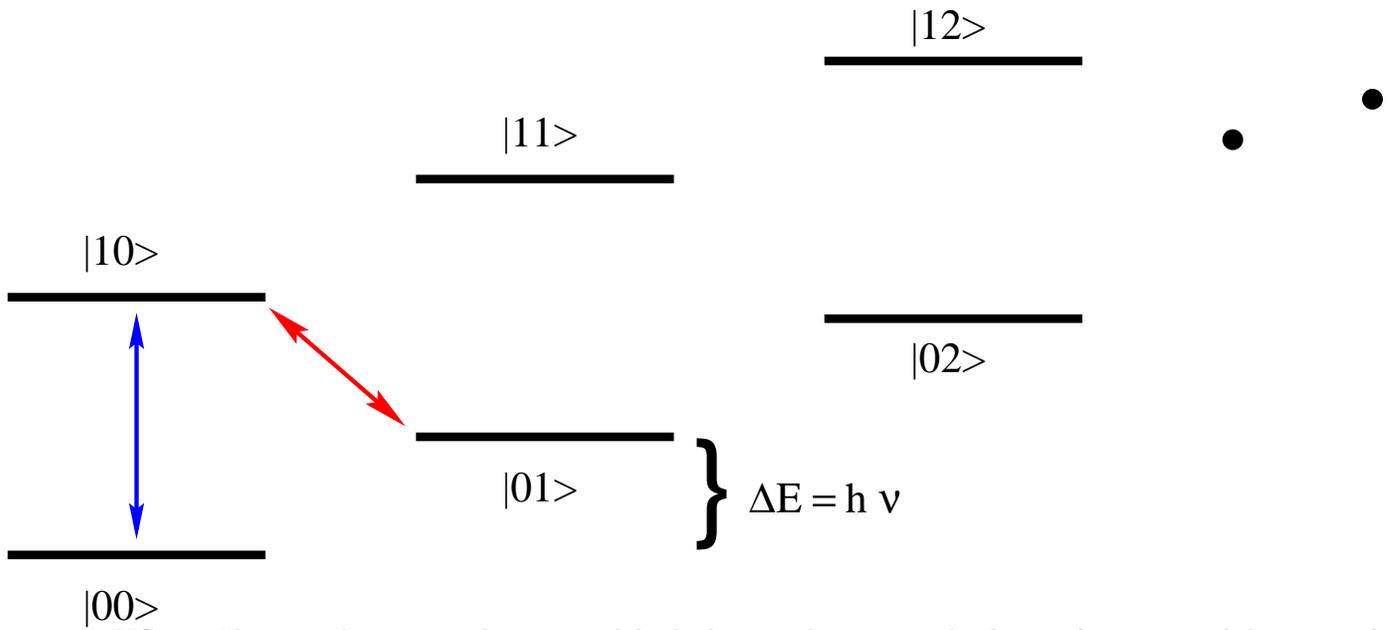}}
\caption{The vertical axis gives the energy while the horizontal axis
gives the degree of excitation of the center-of-mass mode. In $|xy\rangle$
the first number $x$ denotes the internal degree of freedom of the ion,
while the second number $y$ denotes the degree of excitation of the 
center of mass mode. }
\label{ladder}
\end{figure}

\begin{figure}[h]
\centerline{\epsfbox{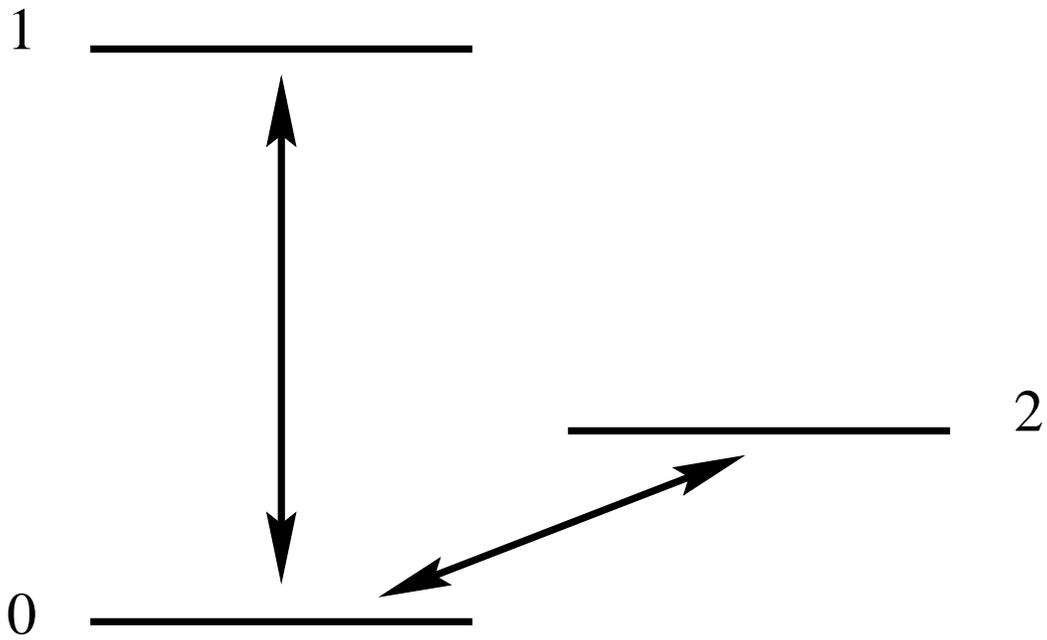}}
\vspace*{0.4cm}
\caption{The energy levels $0$ and $1$ represent the two levels of the qubit. The auxiliary
level $2$ is necessary to implement a CNOT operation in a linear ion trap. }
\label{tlsion}
\end{figure}

\begin{figure}[h]
\centerline{\epsfbox{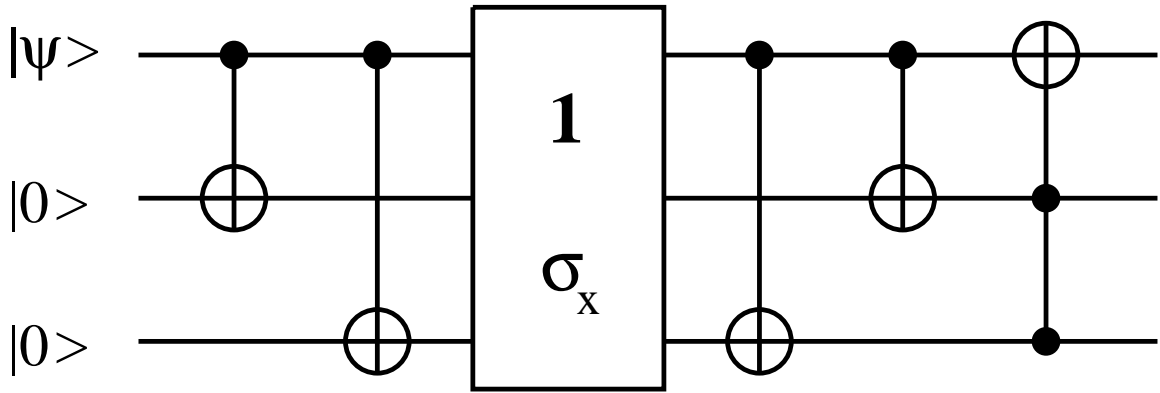}}
\vspace*{0.4cm}
\caption{The quantum network implementing a amplitude error correcting code. 
The initial state $|\psi\rangle$ is encoded into three bits using two
CNOT gates. If at most one bit flip error occurs, indicated by the box, 
then the subsequent decoding and 
error correction using the Toffoli gate restores the state $|\psi\rangle$ 
of the first qubit completely.}
\label{ampl}
\end{figure}

\begin{figure}[h]
\centerline{\epsfbox{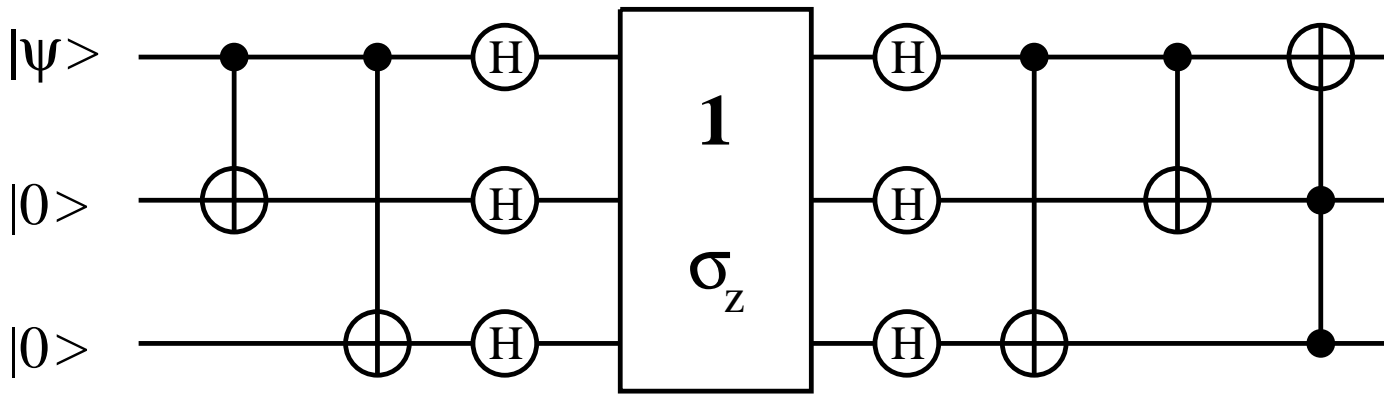}}
\vspace*{0.4cm}
\caption{The quantum network implementing a phase error correcting code. 
The initial state $|\psi\rangle$ is encoded into three bits using two
CNOT gates and subsequent rotation of the basis using Hadamard transformations. 
If at most one phase error occurs, indicated by the box, then the 
subsequent decoding and error correction
using the Toffoli gate restores the state $|\psi\rangle$ of the first qubit completely.}
\label{phase}
\end{figure}

\begin{figure}[h]
\centerline{\epsfbox{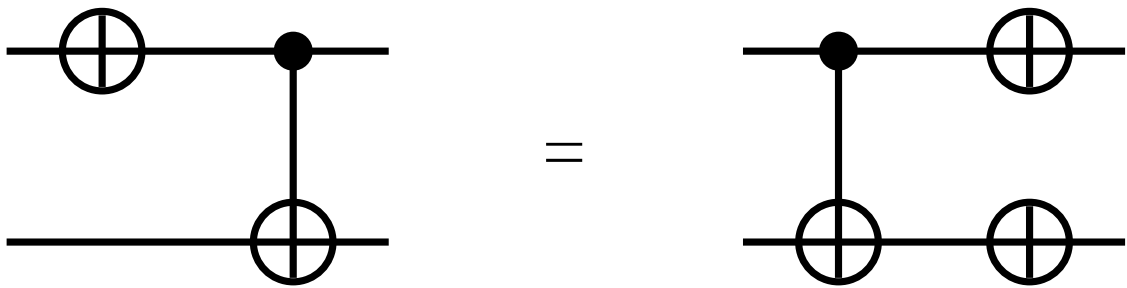}}
\vspace*{0.4cm}
\caption{The two networks are equivalent. On the left hand side the control 
qubit has suffered an amplitude error before the CNOT gate was performed. 
On the right hand side one observes that this is equivalent to have first 
a CNOT gate and then amplitude errors in both qubits. This illustrates that 
errors in one qubit during quantum gate operations can lead to
errors in all of the involved qubits.}
\label{error}
\end{figure}

\end{document}